\documentclass[aps,prb,twocolumn,superscriptaddress,showpacs,amsmath]{revtex4}

\usepackage{bm}
\usepackage{amssymb}
\usepackage{mathrsfs}
\usepackage{graphicx}

\newcommand{\Ref}[1]{Ref.\ \onlinecite{#1}}
\newcommand{\Refs}[1]{Refs.\ \onlinecite{#1}}
\newcommand{\Eq}[1]{Eq.\ (\ref{#1})}
\newcommand{\Eqs}[2]{Eqs.\ (\ref{#1})-(\ref{#2})}

\newcommand{\bfig}[1]{\begin{figure}[{#1}]}
\newcommand{\bwfig}[1]{\begin{figure*}[{#1}]}
\newcommand{\efig}{\end{figure}}  
\newcommand{\ewfig}{\end{figure*}}  
\newcommand{\fig}[2]{\includegraphics*[scale={#1}]{#2}}

\newcommand{\etal}{\textit{et al.}}
\newcommand{\hTc}{high-$T_c$}

\newcommand{\tJ}{$t$-$J$}

\newcommand{\bk}{{\mathbf k}}

\newcommand{\br}{{\mathbf r}}

\newcommand{\bQ}{{\mathbf Q}}

\newcommand{\ket}[1]{|{#1}\rangle}
\newcommand{\bra}[1]{\langle{#1}|}
\newcommand{\braket}[1]{\langle{#1}|{#1}\rangle}
\newcommand{\braketII}[2]{\langle{#1}|{#2}\rangle}
\newcommand{\ketbra}[1]{|{#1}\rangle \langle{#1}|}
\newcommand{\ketbraII}[2]{|{#1}\rangle \langle{#2}|}
\newcommand{\expn}[1]{\langle {#1} \rangle}
\newcommand{\inrm}{\mathrm{in}}
\newcommand{\outrm}{\mathrm{out}}

\newcommand{\ybco}[1]{$\mathrm{YBa_2Cu_3O}_{{#1}}$}
\newcommand{\ybcoeight}{$\mathrm{YBa_2Cu_4O_8}$}
\newcommand{\tbco}{$\mathrm{Tl_2Ba_2CuO_{6+\delta}}$}
\newcommand{\lscox}{$\mathrm{La}_{2-x}\mathrm{Sr}_x\mathrm{CuO_4}$}
\newcommand{\bscco}{$\mathrm{Bi_2Sr_2CaCu_2O_{8+\delta}}$}
\newcommand{\NaCCOC}{$\mathrm{Na}_{x}\mathrm{Ca}_{2-x}\mathrm{CuO_2Cl_2}$}

\begin{document}

\title{Antiferromagnetism and charged vortices in high-$T_c$ superconductors}
\author{Daniel Knapp}
\email[Corresponding author: ]{dan@danielk.ca}
\affiliation{Department of Physics and Astronomy, McMaster University, Hamilton, Ontario, Canada L8S 4M1}
\author{Catherine Kallin}
\affiliation{Department of Physics and Astronomy, McMaster University, Hamilton, Ontario, Canada L8S 4M1}
\author{Amit Ghosal}
\affiliation{Department of Physics, Duke University, North Carolina, 27708-0305 USA}
\author{Sarah Mansour}
\altaffiliation{Current address: Department of Biochemistry, University of Toronto Hospital for Sick Children, 555 University Avenue, Toronto, Ontario, Canada M5G 1X8}
\affiliation{Department of Physics and Astronomy, McMaster University, Hamilton, Ontario, Canada L8S 4M1}

\date{\today}

\begin{abstract}
The effect of the long-range Coulomb interaction on charge accumulation in antiferromagnetic vortices in high-$T_c$ superconductors is studied within a Bogoliubov-de Gennes mean-field model of competing antiferromagnetic and $d$-wave superconducting orders.  Antiferromagnetism is found to be associated with an accumulation of charge in the vortex core, even in the presence of the long-range Coulomb interaction.  The manifestation of $\Pi$-triplet pairing in the presence of coexisting $d$-wave superconductivity and antiferromagnetic order, and the intriguing appearance of one-dimensional stripelike ordering are discussed.  The local density of states in the vortex core is calculated and is found to be in excellent qualitative agreement with experimental data.
\end{abstract}

 %pacs numbers identify the categories of s/c state & magnetic properties respectively

\maketitle

%==========================================================================
\section{Introduction}\label{sec:intro}

The vortex cores of {\hTc} superconductors have been the subject of intense investigation in recent years.  Advances in experimental techniques, especially in the area of scanning tunneling microscopy (STM),\cite{maggio95, renner98, pan00, hoffman02} have allowed unprecedented exploration of the electronic states at the atomic scale.  Local information on vortex core states has also been provided by high-field nuclear magnetic resonance (NMR) experiments,\cite{curro00, mitrovic01, kakuyanagi02, mitrovic03, kakuyanagi03} which use inhomogeneity in the magnetic field to extract a spatially resolved relaxation rate.

Among the principal results of these experiments is the suppression of the low-energy density of states in the vortex core,\cite{mitrovic01, kakuyanagi02} with either an absence, or possibly a strong splitting of the zero-bias conductance peak (ZBCP),\cite{maggio95, renner98, pan00} indicating some form of ordering in the vortex core.  STM results also show the presence of low-energy vortex core states which extend beyond the vortex core radius,\cite{maggio95, pan00, hoffman02} as defined by the superconducting coherence length $\xi$. These extended vortex-induced states were further observed to form a localized checkerboard pattern along the CuO bonds with a period near $4.3a_0$ (\Ref{hoffman02}).  A similar checkerboard pattern with a period close to $4.5~ a_0$ has also been seen in more recent STM experiments in zero field, both in the pseudogap state\cite{vershinin04} and in the superconducting state\cite{howald03, davis04} of {\bscco} (BSCCO).  These observations are corroborated by STM experiments on the superconductor {\NaCCOC} (Na-CCOC) which show strong ordering with a period of $4 a_0$ in a pseudogaplike state.\cite{hanaguri04}  It has been proposed that this checkerboard pattern is due to the formation of a pair-density wave of phase-incoherent Cooper pairs;\cite{zhang02, zhang04, tesanovic04} however, this interpretation remains controversial and the existence of other orderings, such as dynamically fluctuating stripes,\cite{kivelson03, norman04} spin-density wave ordering,\cite{sachdev02_rapid} or a particle-hole charge-density wave,\cite{fu04} remains a likelihood.  Indeed, neutron-scattering experiments have reported the existence of a periodic modulation of antiferromagnetic correlations in zero field.\cite{keimer99, keimer02, keimer04_nature, keimer04, tranquada04, hayden04, christensen04, stock04_prb, stock04} 
 
Many experiments have also reported observations of field-induced magnetic order in the {\hTc} superconductors.   It has been noted that the periodicity of the charge ordering observed in STM (\Ref{hoffman02}) could be consistent with inelastic neutron-scattering experiments on optimally doped {\lscox} ($x = 0.163$) that show a field-induced signal in the low-frequency spin-fluctuation spectrum, near the $(\pi,\pi)$ point in reciprocal space, suggesting the existence of a fluctuating spin-density wave with a periodicity close to $8 a_0$ along the CuO bond directions and a correlation length $\ell > 20$ lattice spacings.\cite{lake01}  More local measurements using spatially resolved NMR also provide strong evidence for the presence of antiferromagnetic order in the vortex cores of near-optimally doped {\ybco{7-\delta}} (\Ref{mitrovic03}) and {\tbco} (\Ref{kakuyanagi03}).  Muon-spin resonance measurements of the magnetic field distribution of underdoped {\ybco{6.50}} reveal structure in the high-field tail that is not seen in optimally doped {\ybco{6.95}} and is consistent with static antiferromagnetism.\cite{miller02}    Elastic neutron-scattering has shown a strong enhancement of incommensurate static antiferromagnetism in an applied magnetic field in underdoped {\lscox} with $x = 0.10$ (\Ref{lake02}) and with $x = 0.12$ (\Ref{katano00}).  The antiferromagnetism (AFM) is observed to have a very long correlation length {$> 400$ {\AA}} showing that AFM and superconductivity (SC) coexist in the bulk.\cite{lake02}  Similar results are seen in elastic neutron-scattering experiments on $\mathrm{La_2CuO}_{4+y}$ (\Ref{khaykovich02}).  The picture that emerges from these various measurements is that static or dynamic antiferromagnetic order is induced, or at least strongly enhanced, by an applied magnetic field.  This antiferromagnetic order appears to be nucleated at the vortex cores, and extends into the bulk of the superconductor with a correlation length that is much longer than the superconducting coherence length.

Charge-density modulations induced at vortex cores by the presence of antiferromagnetic or spin-density-wave order have been predicted within various theoretical approaches.\cite{arovas97, sachdev01, sachdev02, sachdev02_science, sachdev02_rapid, ghosal02, ting02_prl, zhu02}  In partial agreement with the experimental results discussed above, common features of many of these calculations include an antiferromagnetic or spin-density-wave order that is nucleated at the vortex core (where $d$-wave superconductivity is most strongly suppressed) and which extends well beyond the core region into the bulk of the superconductor.  This antiferromagnetic order is accompanied by a local modulation in the charge density that reaches a maximum at the centre of the vortex core and is seen to approach half filling in some cases.\cite{ghosal02, ting02_prl}  The vortex core charge is predicted to have a strong doping dependence.\cite{ting02_prl} In the overdoped region of the phase diagram the vortex cores are not antiferromagnetic, and the difference between the chemical potentials of the normal core and the surrounding superconducting region leads to a small decrease in the charge density at the core.\cite{khomskii95, blatter96}  In this case the vortices are positively charged.  As the doping is decreased, the vortex cores become antiferromagnetic and expel holes into the bulk of the superconductor, causing  the vortices to become negatively charged.  

Since there is no reason to expect pair-density-wave ordering to push the local charge density toward half filling, the presence of negatively charged vortices in the underdoped cuprates could be a distinguishing feature of field-induced antiferromagnetic ordering. However, the ability of this induced charge to survive the effects of the long-range Coulomb (LRC) interaction has not been established.  In order to determine whether negatively charged vortices can persist as a robust prediction of a theory containing antiferromagnetically ordered vortices, we have set out to include the long-range Coulomb interaction in a self-consistent mean-field theory of the cuprates that contains both antiferromagnetic and superconducting order parameters.  The self-consistent effect of the long-range Coulomb interaction on the doping and magnetic field dependence of the antiferromagnetic order and vortex-core charge is calculated.  Our results suggest that an antiferromagnetic vortex core is \textit{always} associated with an accumulation of electrons in the core region, even in the presence of the long-range Coulomb interaction.  However, both the AFM order and the associated vortex core charge are weakened with increased LRC interaction strength, and eventually vanish \textit{simultaneously} above a critical value.  
We further find that the local density of states (LDOS) in the vortex core is modified by the long-range Coulomb interaction in such a way that the essential features of STM data on YBCO (\Ref{maggio95}) are captured within a simplified model of competing AFM and $d$SC orders. We also study the manifestation of $\Pi$-triplet pairing in the presence of coexisting $d$SC and AFM order, and discuss the intriguing appearance of one-dimensional stripelike ordering.

%==========================================================================
\section{Model and Method} \label{sec:model}

We describe a two-dimensional $d$-wave superconductor in an external magnetic field by the {\tJ} Hamiltonian with an additional long-range Coulomb interaction
%Long range Coulomb
\begin{equation}
\mathcal{H}_{\mathrm{C}} = \frac{V}{2}\sum_{i \ne j} \frac{n_i n_j}{|\mathbf{r}_{ij}|},
\label{eq:ham_lrc}
\end{equation}
where $V=e^2/ \epsilon_h a_0^2$ with dielectric constant $\epsilon_h$, and $|\mathbf{r}_{ij}|$ is in units of the lattice constant $a_0$.  The LRC interaction is added to the Hamiltonian in order to study the competing effects of the local antiferromagnetic order in the vortex cores and the Coulomb repulsion between electrons.
\cite{ting04}${}^,$
\footnote{Since we are mainly interested in the effects of long-range screening, we neglect the on-site Coulomb repulsion.  Including the on-site term would lead to additional AFM and triplet-pairing order parameters and would complicate a direct comparison with $V=0$ results.  We note that mean-field solutions of the {\tJ} model at $V=0$ give the same qualitative results as are seen in mean-field solutions of the extended Hubbard model (see for example, \Ref{ting04}) which include an on-site Coulomb repulsion.}
We start with the mean-field model of Ghosal {\etal}\cite{ghosal02} and include the effects of the long-range Coulomb interaction over all sites for the Hartree shift (which includes the density-density interaction) and at the nearest-neighbor level for all other order parameters.  It is important to note that screening effects will be accounted for by self-consistent calculation of the Hartree shift---which allows for rearrangement of the charge density---as in density functional theory (a similar approach has been taken in a zero-field study of the $t$-$J$ model by Arrigoni {\etal}\cite{arrigoni02}).  With the LRC interaction incorporated in this way, the effective Hamiltonian of \Ref{ghosal02} becomes
%Mean field Hamiltonian
\begin{eqnarray}
\mathcal{H}_\mathrm{eff}= &-&\sum_{i,\delta,\sigma} 
\left(t + \tau_{i,\sigma}^{\delta}\right)
e^{i\phi_{i}^{\delta}} c_{i, \sigma}^{\dag} c_{i+\delta,\sigma} \nonumber \\
&-&\sum_{i,\sigma} 
\left( \mu - \eta_i + \sigma e^{i \mathbf{Q \cdot r}_i} \sum_{\delta} m_{i+\delta} \right) n_{i,\sigma}\nonumber \\
&+& \sum_{i,\delta,\sigma}
\left( \Delta_{i, \sigma}^{\delta} c_{i, \sigma}^{\dag} c_{i+\delta, -\sigma}^{\dag} + {\rm H.c.}\right),
\label{eq:heff}
\end{eqnarray}
where $\tau_{i,\sigma}^{\delta}$ is the Fock shift that renormalizes the hopping amplitude $t$ (we set $t=1$ in this paper), and $\eta_i$ is the Hartree shift that renormalizes the chemical potential $\mu$.  Here $\delta = \pm \hat x, \pm \hat y$, the four nearest neighbors of any site, and $\phi_i^{\delta} = (\pi/\Phi_0) \int_{i}^{i+\delta} \mathbf{A}(\br) \cdot d \br$.  The local $d$SC ($\Delta_{i, \sigma}^{\delta}$) and AFM ($m_i$) order parameters satisfy the following self-consistency relations:
%self-consistency relations for delta and AFM
\begin{equation}
\label{eq:sc1}
\Delta_{i, \sigma}^{\delta} = \left\{
\frac{J}{4} \langle c_{i+\delta,-\sigma} c_{i, \sigma} \rangle 
-\left( \frac{J}{4}-\frac{V}{2} \right) \langle c_{i+\delta, \sigma} c_{i, -\sigma} \rangle
\right\}
\end{equation}
and
\begin{equation}
\label{eq:sc2}
m_i = \frac{J}{4} \left( \langle n_{i\uparrow} \rangle -
\langle n_{i\downarrow} \rangle \right) e^{i \mathbf{Q \cdot r}_i},
\end{equation}
where $\mathbf{Q}=(\pi,\pi)$ is the antiferromagnetic wave vector.  The Fock shift 
%fock shift
\begin{equation}
\tau_{i,\sigma}^{\delta} = \frac{1}{2}\left\{ \frac{J}{4} \langle c_{i+\delta,-\sigma}^{\dag}
c_{i,-\sigma} \rangle
+ \frac{V}{2} \langle c_{i+\delta,\sigma}^{\dag}
c_{i,\sigma} \rangle \right\}  e^{-i \phi_i^{\delta}}
+ \mathrm{H.c.}
\label{eq:sc3}
\end{equation}
is assumed to have no dependence on the direction of hopping across a given bond, and is allowed to renormalize only the magnitude and not the phase of the hopping energy $t$.  This is consistent with the assumption employed here that the magnetic field is uniform throughout the sample.  The Hartree shift
%Hartree shift
\begin{equation}
\eta_i = - \frac{J}{4}\sum_{\delta}\langle
n_{i+\delta} \rangle 
+ V \sum_{j \in \mathrm{u.c.}  } \alpha_{i,j} \langle n_j \rangle,
\label{eq:sc4}
\end{equation}
where the sum in $j$ is over all sites in a magnetic unit cell, contains the density-density part of the long-range Coulomb interaction.  The constants $\alpha_{i,j}$ are calculated by the method of Ewald summation with the assumption that the magnetic unit cell is periodically repeated in all three directions in space, separated by a distance in the perpendicular direction of $c \sim 3 a_0$, and surrounded by a uniform, neutralizing background charge with a static dielectric constant $\epsilon_h$. In this paper, we report on results for $V$ from 0 to 0.35.  For $t \sim 0.5$ eV, this corresponds to a static dielectric constant for the bound charges from $\epsilon_h \to \infty$ down to  $\epsilon_h \sim 20$.   Since we are working at high fields, we assume a square vortex lattice\cite{gilardi02, brown04} with nearest-neighbor vortices along the Cu-O bond directions: our results, which focus on the vortex core structure, do not depend on this choice.

In order to diagonalize the effective Hamiltonian, we make a Bogoliubov-de Gennes (BdG) transformation and block diagonalize the resulting BdG Hamiltonian by exploiting the magnetic-translational symmetry of the problem.  Starting with an initial guess for all \textit{local} variables and for the value of $\mu$, we solve for all eigenvalues and eigenvectors of the BdG matrix for each magnetic wave vector $\bk$. Results in this paper are reported for $24 \times 12$ wave vectors. Note that the time-reversal symmetry of the BdG Hamiltonian,\cite{degennes66} $E_{\sigma}(\bk) \to -E_{-\sigma}(-\bk)$, allows us to find the eigenvalues and eigenvectors of the $\bk$-dependent blocks of the BdG matrix in $(\bk, -\bk)$ pairs
\begin{equation}
\begin{pmatrix}
u_{i,\sigma}(\bk) \cr v_{i,-\sigma}(\bk)
\end{pmatrix}
 \to
\begin{pmatrix}
-v_{i,-\sigma}^{*}(-\bk) \cr u_{i,\sigma}^*(-\bk)
\end{pmatrix}.
\end{equation}
The eigenvalues and eigenvectors are used to recalculate all of the local variables using the self-consistency equations defined above [\Eqs{eq:sc1}{eq:sc4}]. This iterative process is repeated until the largest change in all site-dependent variables and in the average charge density (which is controlled by tuning the chemical potential $\mu$) is less than $10^{-5}$.  The addition of the long-range Coulomb interaction destabilizes this iterative procedure, and we found it necessary to develop a modified Broyden's method to achieve convergence in the Hartree shift.  The details of the convergence method are discussed in appendix \ref{sec:convergence}.

\section{Results} 
\label{sec:results}

%=================================================================
\subsection{Effect of the long-range Coulomb interaction on the vortex core charge}
\label{ssec:lrc1}

In the absence of the long-range Coulomb interaction the suppression of $d$-wave superconductivity at the vortex cores leads to the emergence of antiferromagnetic order with a correlation length $\ell_m$ that is much longer than the superconducting coherence length $\xi$, for the range of parameters that is appropriate to the cuprates.  The antiferromagnetism is strongest at the center of the vortex core, and creates a non uniform charge density through the expulsion of holes from the vortex core into the bulk of the superconductor.  In this paper, we focus our attention on the added effects of the long-range Coulomb interaction on this charge ordering.  

As one would expect, the principal effect of including the Coulomb repulsion is to reduce the magnitude of the peak charge density at the center of the vortex core.  As can be seen in Fig.\ \ref{fig:den_U},
\bwfig{p}
\fig{0.6}{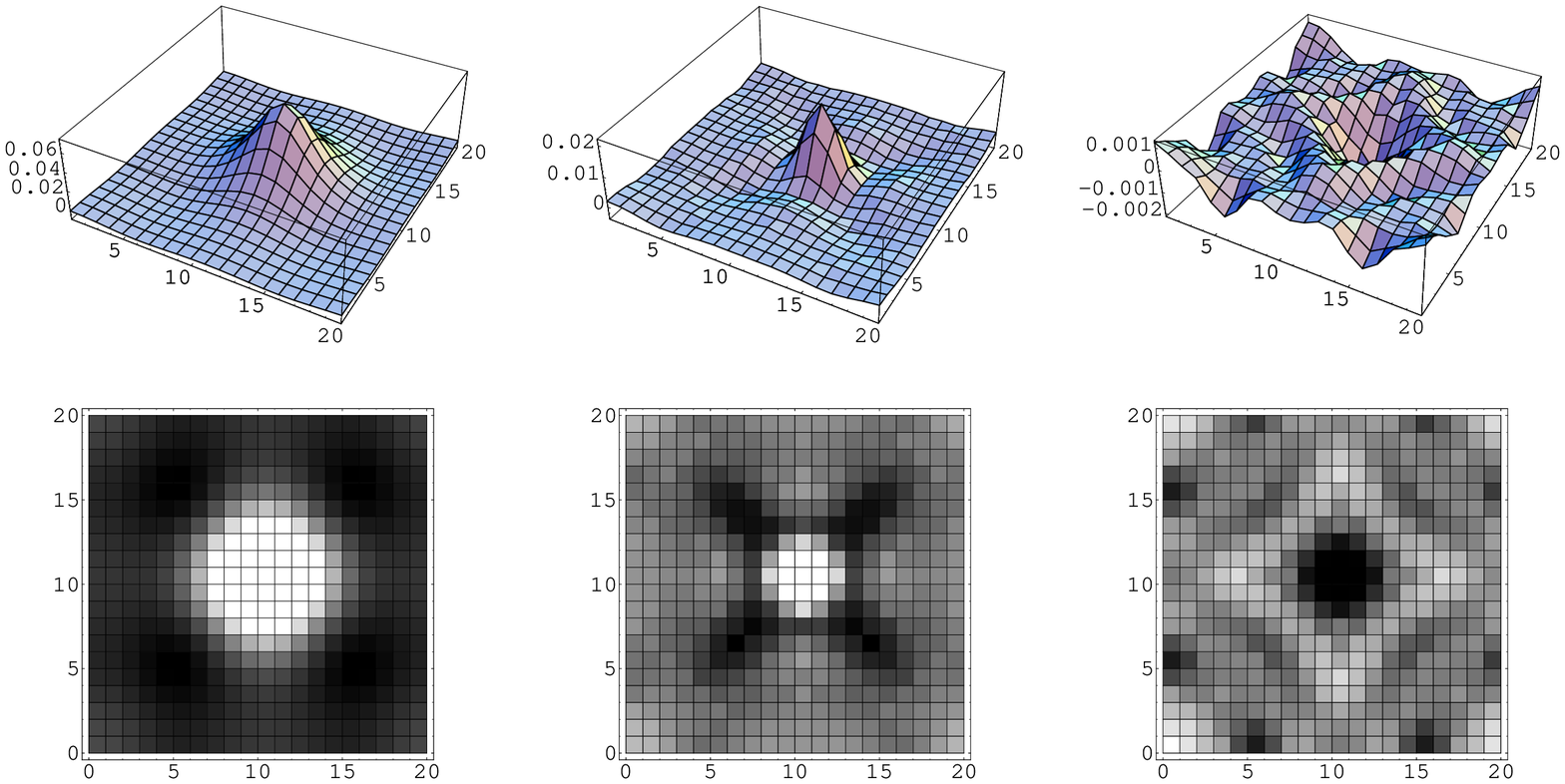}
\caption{\label{fig:den_U} From left to right, electron number density relative to the average density $\delta \expn{n_i} = \expn{n_i} - 0.875$, for $V=0$, 0.2, and 0.35 on one half of a $20 \times 40$ unit cell (the other half is equivalent by symmetry) for an interaction strength $J=1.15$.  For $V=0.2$ a region of reduced electron density screens the peak in the core.  At $V=0.35$ the vortex core charge and the screening charge have changed sign, and modulations in the electron density have a much smaller amplitude (on the order of  $10^{-3}$ electrons per site).  The bottom row shows two-dimensional (2D) density plots of the 3D plots in the top row.}
\ewfig
which plots the non uniform charge density $\delta \expn{n_i} = \expn{n_i} - n_{\mathrm{ave}}$, the large non-uniform electron density in the vortex core is first reduced and then eliminated as the LRC strength is increased from $V=0$ to $V=0.35$.  Initially, the effect of the LRC interaction is to sharpen the charge density profile and to induce a screening region of reduced charge density immediately around the vortex core.  This screening region separates the charge density profile into two distinct regions:  inside the core AFM order governs the charge density, pushing it toward half filling; outside the core AFM no longer controls the charge density and there is a weak dip in the charge density along the nodal directions beyond the screening region, due to the effects of $d$SC order.  

Interestingly, the shape of the antiferromagnetic ordering (see Fig.\ \ref{fig:mag_U}) 
\bwfig{p}
\fig{0.6}{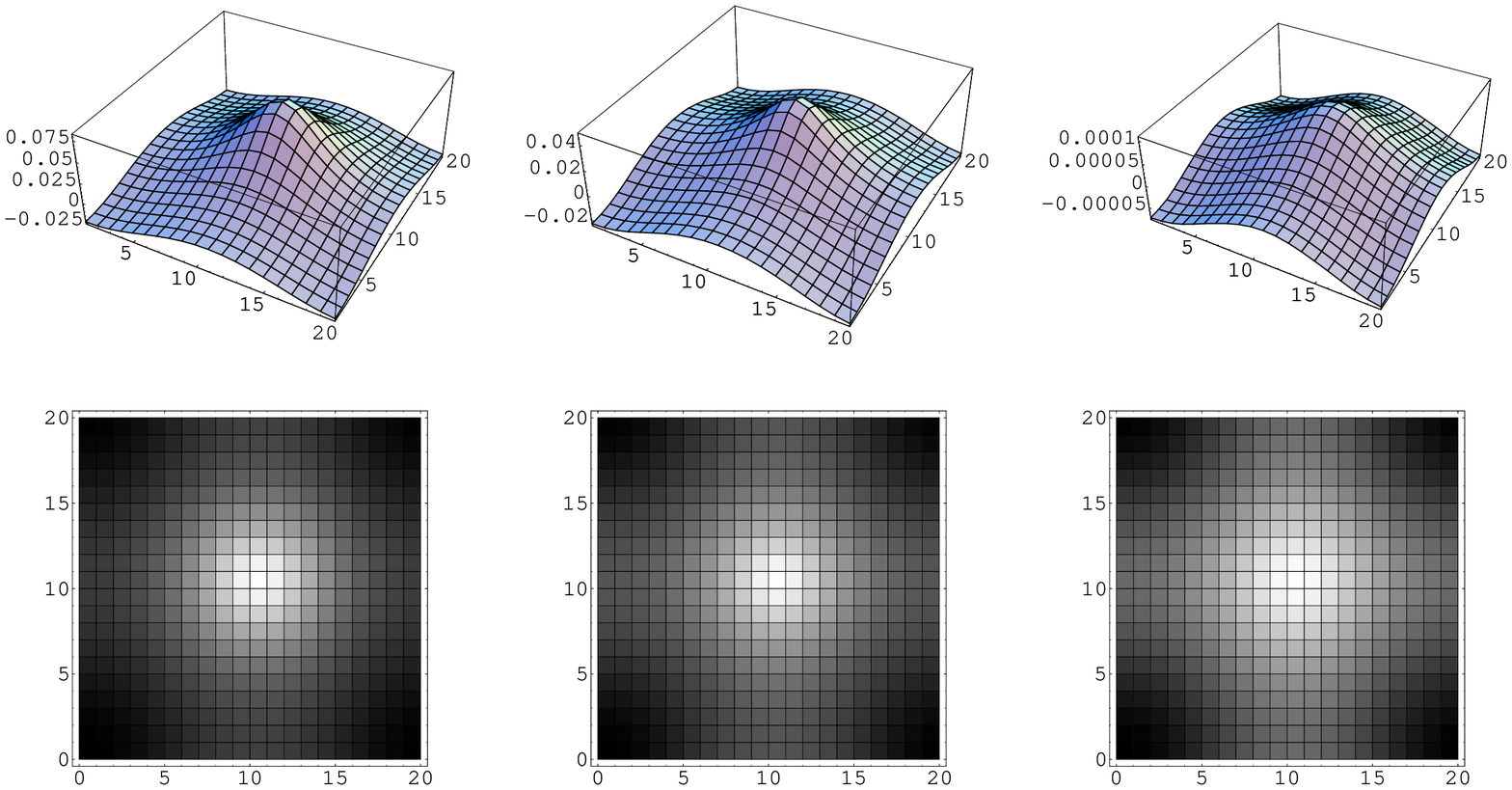}
\caption{\label{fig:mag_U} From left to right, the AFM order $m_i$ (in units of $t$) for $V=0$, 0.2, and 0.35 on one half of a $20 \times 40$ unit cell, with $J=1.15$ and $n_{\mathrm{ave}}=0.875$.  With $V=0.2$ the magnitude of the AFM has been approximately halved and at $V=0.35$, where the vortex core is no longer negatively charged, the AFM order is negligible.  The shape of the AFM order is unaffected by the dramatic changes in the structure of the electron density induced by the LRC interaction.  The bottom row shows 2D density plots of the 3D plots in the top row.}
\ewfig
is not affected by these dramatic changes in the short-range structure of the charge density.  This is reflective of the long correlation length of the AFM order.  However, the magnitude of the AFM order is reduced along with that of the non uniform charge density. As the interaction strength is further increased to $V=0.35$, the charge density modulations are greatly weakened, with a small amplitude on the order of $10^{-3}$ electrons per site.   At this point, the AFM order at the vortex core has been completely destroyed ($|m| \sim 10^{-4}$) by the reduction in the local charge density.  In the absence of AFM order, the vortex core becomes weakly positively charged due to the gap variation between the ``normal'' core and the surrounding superconductor.\cite{khomskii95, blatter96}  The charge density profile at $V=0.35$ is equivalent to that seen in more highly doped systems where AFM order is absent even at $V=0$ (discussed below in Sec.\ \ref{ssec:doping}).  

The $d$-wave superconductivity ($d$SC) is also suppressed by the Coulomb repulsion, which weakens the nearest-neighbor pairing attraction [as can be seen in the second part of Eq.\ (\ref{eq:sc1})].  Fig.\ \ref{fig:delta_U}
\bwfig{p}
\fig{0.6}{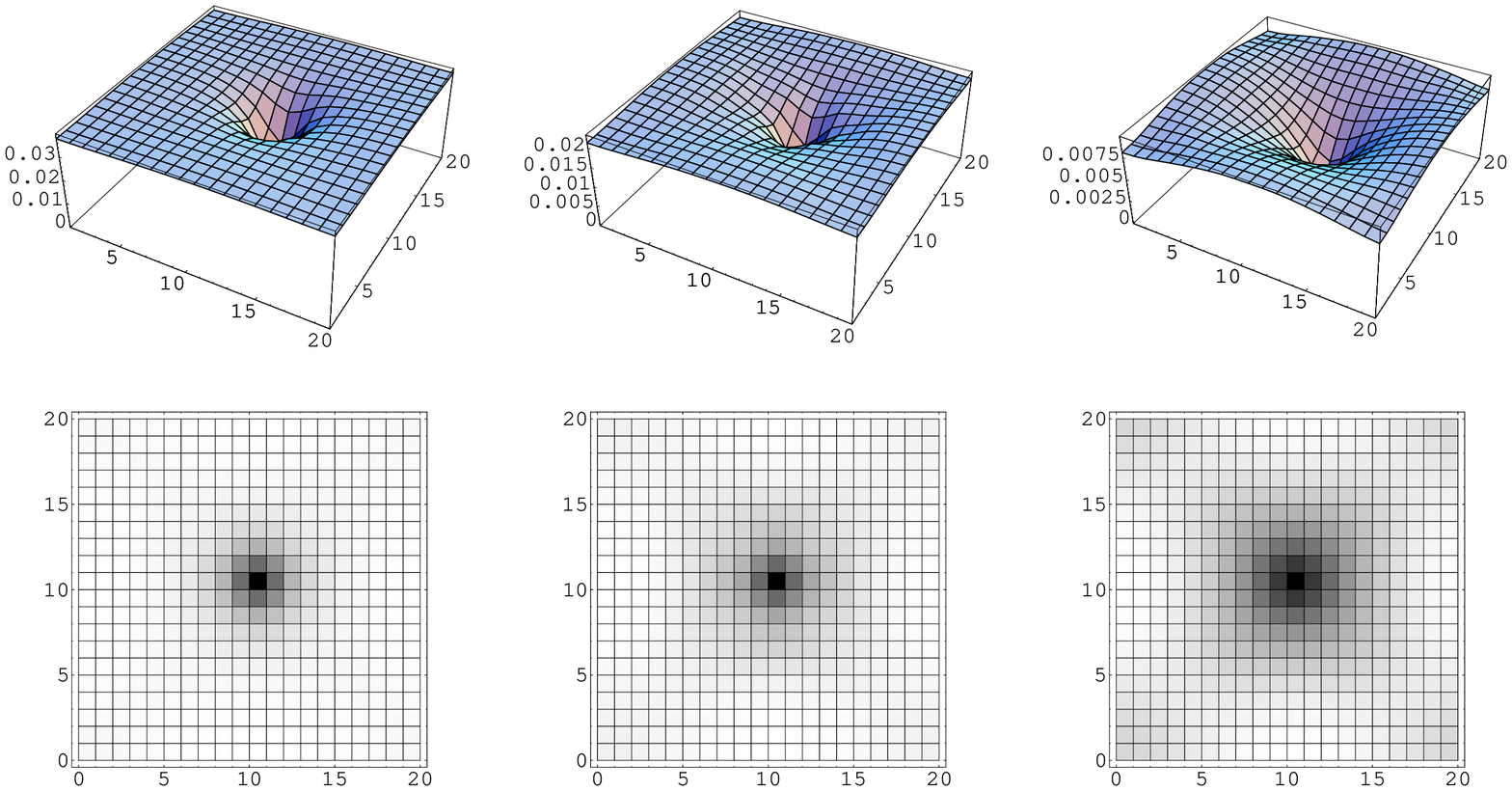}
\caption{\label{fig:delta_U} From left to right, the magnitude of the $d$SC order $|\Delta_i|$ (in units of $t$) for $V=0$, 0.2, and 0.35 on one half of a $20 \times 40$ unit cell, with $J=1.15$ and $n_{\mathrm{ave}}=0.875$.  The structure of $|\Delta_i|$ is unchanged by the LRC, but the magnitude is gradually reduced due to weakening of the nearest-neighbor attraction that generates the $d$SC pairing.  The bottom row shows 2D density plots of the 3D plots in the top row.  One can see that the coherence length $\xi$ increases as the strength of the $d$-wave pairing is reduced.}
\ewfig
shows the evolution of the $d$SC order as the Coulomb interaction is increased from $V=0$ to $V=0.35$.   The overall shape of the $d$SC order does not change; but, the coherence length gradually increases as the magnitude is reduced.   Note that it is the site-dependent gauge-invariant gap operator, defined as 
\begin{equation}
\Delta_i = \frac{1}{8} \sum_{\delta, \sigma} (-1)^{\delta_y}e^{-i\phi_{i}^{\delta}} \sigma \Delta_{i, \sigma}^{\delta},
\end{equation}
that is shown.

It is clear that the survival of antiferromagnetic order at the vortex core is dependent upon the existence of a local charge density that is closer to half filling than the average density.  This dependence is evident in Fig.\ \ref{fig:mag_den_delta_U},
\bfig{p}
\fig{0.48}{figure4.eps}
\caption{\label{fig:mag_den_delta_U} Peak values of $d$SC $|\Delta_0|$ ($\Box$) far from the vortex core, and AFM $|m_{\mathrm{core}}|$ ($\circ$) and electron density $\delta \expn{n_{\mathrm{core}}}$ ($\bigtriangleup$) at the center of the vortex core as a function of increased LRC strength $V$, for a $20 \times 40$ unit cell with $J=1.15$ and $n_{\mathrm{ave}}=0.875$.  All quantities are plotted relative to their $V=0$ values.  Between $V=0.3$ and $V=0.35$, the density of electrons in the core crosses over from positive to negative and the AFM order disappears.   Lines are a guide to the eye. }
\efig
which shows the effects of the long-range Coulomb interaction on the strength of the $d$SC, on the AFM order, and on the peak charge density in the vortex core.   The AFM decreases gradually until $V > 0.2$, where it begins to rapidly fall off. Both the antiferromagnetism and the charged vortex core disappear between $V=0.3$ and $V=0.35$.  The $d$SC order $|\Delta_0|$ decreases linearly at first and then gradually starts to fall off more quickly with $V$, which is consistent with the reduction of the pairing interaction strength in the second part of \Eq{eq:sc1}, and is able to survive at larger values of $V$ where the AFM and the charge order are destroyed.  The existence of antiferromagnetic order clearly implies, and is dependent upon, the existence of negatively charged vortex cores.

%======================================================================
\subsection{Magnetic field dependence of antiferromagnetic order and core charge density}
\label{ssec:size}

In an attempt to better characterize the effect of the long-range Coulomb interaction on the vortex core charge we have also studied the dependence of the AFM order, and its accompanying charge density order, on the size of the unit cell ($H = 2 \Phi_0 / N_x N_y a_0^2 \approx 34$ T for a $20 \times 40$ unit cell).  At $V=0$, the AFM increases and the charge density moves closer to half filling with increased unit cell size.   In a larger unit cell the AFM order, which has a correlation length $\ell_m >> \xi$, is able to reach larger values by spreading out its variations over longer distances.
Fig.\ \ref{fig:af_den_size} 
\bfig{p}
\fig{0.48}{figure5.eps}
\caption{\label{fig:af_den_size} Antiferromagnetic order $m_{\mathrm{core}}$ (dotted lines), in units of $t$, and electron number density $\delta \expn{n_{\mathrm{core}}}$ (solid lines) at the center of the vortex core as a function of unit cell size $L \times 2 L$ for LRC strengths of $V=0$ ($\circ$), $V=0.15$ ($\Box$) and $V=0.25$ ($\Diamond$), with  $J=1.15$ and $n_{\mathrm{ave}}=0.875$.  When $V=0$ the density at the core increases toward half filling (0.125 in this plot) with increasing unit cell size.  For $V>0$ the dependence becomes non-monotonic with the peak charge density occurring in the $22 \times 44$ unit cell.  The anomalous increase in the AFM at $V=0.25$ for $L=24$, 26 is due to the development of a 1D anisotropy discussed in Sec.\ \ref{ssec:stripes}.  Lines are a guide to the eye.}
\efig
shows that the LRC interaction has a dramatic effect on this relationship: when $V>0$, the AFM and the charge density at the center of the vortex core initially increase, but then start to decrease with increasing unit-cell size above a unit-cell size of $22 \times 44$.

As discussed above, the shape of the AFM order does not change with increased $V$, despite dramatic changes in the structure of the charge density.  The observed reduction in AFM order in larger unit cells is likely to due to the fact that the peak in the AFM is spread out over large distances.  At $V=0$ this is to the benefit of the AFM; however, when the LRC interaction is turned on the charge density peak is sharpened, and a hole rich region of screening charge forms around it.  As the unit cell gets larger, more of this hole-rich region will lie underneath the central peak in the AFM order.  We suggest that the weakening of the AFM observed in larger unit cells is due to this lowering of the `average' charge density under the peak in the AFM order.  
        
Taken to its limit, this argument suggests that AFM will eventually disappear as the magnetic field is lowered, \textit{if} it maintains the spatial structure commensurate with that of the supercurrent that is seen in Fig.\ \ref{fig:mag_U}.  It is, however, more likely that the spatial structure of the AFM will change at some point as the unit cell size is increased beyond the AFM correlation length $\ell_m$, preserving the AFM vortex core.  Indeed, such an effect is discussed below in Sec.\ \ref{ssec:stripes}.  It is also important to note that AFM order with shorter-range 2D spatial structure that is not tied to the unit cell size, as seen in other work,\cite{franz02, ting02, zhu02, sachdev01, sachdev02_rapid, sachdev02} could be accessed within our model by tuning of the model parameters and introducing a next-nearest-neighbor hopping.\cite{zhu02, ting02}  Such shorter-range ordering may well be less perturbed by the screening effect of the long-range Coulomb interaction at low magnetic fields.

%======================================================================
\subsection{Doping dependence}
\label{ssec:doping}

We have also studied the dependence of the vortex core charge on the average density of electrons, or the doping.  For $V=0$, our results are in agreement with those of Chen {\etal} who observed positively charged vortices with negligible AFM order for larger values of the hole density, with a transition to negatively charged AFM vortices as the average density approaches half filling.\cite{ting02_prl}  For a unit cell of $20 \times 40$ sites with two vortices this transition from positive to negative vortices occurs near an average density of $n_{\mathrm{ave}}=0.855$, at $V=0$ (see Fig.\ \ref{fig:mag_den_doping}).   
\bfig{p}
\fig{0.48}{figure6.eps}
\caption{\label{fig:mag_den_doping} Antiferromagnetic order $m_{\mathrm{core}}$ (dotted lines) in units of $t$, and electron number density $\delta \expn{n_{\mathrm{core}}}$ (solid lines) at the center of the vortex core as a function of the average density $n_{\mathrm{ave}}$ for LRC strengths of $V=0$ ($\circ$), $V=0.15$ ($\Box$) and $V=0.25$ ($\Diamond$) for a $20 \times 40$ unit cell with $J=1.15$.  The region containing AFM and negatively charged vortices shrinks slowly at first and then rapidly toward half filling with increasing $V$. Lines are a guide to the eye. }
\efig
 The chief effect of increasing $V$ is to push this transition closer to half filling and to reduce the magnitudes of both the AFM order and the vortex core charge. The AFM region shrinks gradually at first and then rapidly with increasing $V$.  This is consistent with the behavior seen in Fig.\ \ref{fig:mag_den_delta_U}, where the AFM at the vortex core is seen to decrease more rapidly at larger $V$.  The transition from positively to negatively charged vortices occurs when antiferromagnetism begins to be nucleated in the vortex cores.

%======================================================================
\subsection{$\Pi$-triplet pairing}
\label{ssec:triplet}

It has been shown that $\Pi$-triplet order\cite{zhang97, murakami98, kyung00, ghosal02} is able to develop self-consistently in the presence of coexisting $d$SC and AFM order, even in the absence of any interaction which generates it directly.  The local $\Pi$-triplet order parameter
\begin{equation}
\label{eq:triplet}
\Pi_i^{\delta} = \frac{J}{4}\{ \expn{c_{i+\delta, \downarrow}c_{i,\uparrow}} - \expn{c_{i,\downarrow} c_{i+\delta,\uparrow}}  \} e^{i \bQ \cdot \br_i},
\end{equation}
where $\bQ = (\pi, \pi)$, has a triplet pairing amplitude in the $S_z = 0$ channel.  The prefactor of $J/4$ is introduced for the sake of comparison with the other order parameters.  We find that the shape of the $\Pi$-triplet order does not change with increasing $V$: within the vortex core, $|\Pi_i|$ rises rapidly from zero to its maximum value within a superconducting coherence length $\xi$; outside of the vortex core $|\Delta_i|$ is approximately uniform, and the shape of $|\Pi_i|$ follows that of the AFM order, falling with increasing distance from the core to zero at the lines of zero supercurrent before rising again to smaller peaks in the corners of the unit cell.   The magnitude of the $\Pi$-triplet order decreases with increasing LRC, along with that of the AFM order, and falls below $10^{-6}t$ when $V$ reaches 0.35. 

The $\Pi$-triplet order parameter was introduced by Zhang\cite{zhang97} in the context of SO(5) theory, which proposes a unified theory of AFM and $d$SC order for the cuprates.  It was argued on general symmetry grounds that a theory containing $\Delta_i$, $\Pi_i$, and $m_i$ orders will satisfy the relation 
\begin{equation}
\label{eq:pisymmetry}
\tilde{\Pi}^*_i \tilde{\Delta}_i = - \tilde{m}_i (1 - \expn{n_i}),
\end{equation}
where the variables without the ``tilde'' are those of Eqs.\ (\ref{eq:sc1}), (\ref{eq:sc2}), and (\ref{eq:triplet}) without the prefactors indicating the interaction strength. At $V=0$ this symmetry relation is satisfied within 5\% on a site-by-site basis.  When $V>0$ we find that \Eq{eq:pisymmetry} is still satisfied within 10--20 \% over most of the unit cell, except in the vortex core where the charge density is pushed away from half filling such that $(1-\expn{n_{\mathrm{core}}})$ no longer approaches zero.
In the presence of the LRC interaction, $\Pi$-triplet pairing is observed, but the symmetry relation connecting $\Pi_i$, $\Delta_i$, $m_i$ and the local hole density is weakened.

%======================================================================
\subsection{One-dimensional anisotropy}
\label{ssec:stripes}

When the exchange interaction strength $J$ is increased above 1.15, the AFM and the charge density spontaneously develop a slight one-dimensional anisotropy at $V=0$: both the AFM and charge orders extend to a greater distance along the $x$ or $y$ direction (parallel to the Cu-O bonds).  We have checked the validity of this behavior by repeating the calculations on a square $40 \times 40$ site unit cell containing four vortices, and by comparing the ground state energies of systems with $x$- and $y$-oriented anisotropy in both rectangular (two vortices) and square (four vortices) unit cells.  All results were found to be equivalent within the limits of numerical accuracy.   Similar behaviour has been seen in other BdG studies of the vortex state based on mean-field solutions of the extended Hubbard model with added nearest-neighbor pairing.\cite{ting02, ichmata04} 

We have found that this one-dimensional anisotropy is strongly enhanced by the long-range Coulomb interaction. At $V=0$ (not shown) the AFM and the charge density orders have an elliptical shape.  At $V=0.35$, as shown in Fig.\ \ref{fig:stripes},
\bwfig{p}
\fig{0.6}{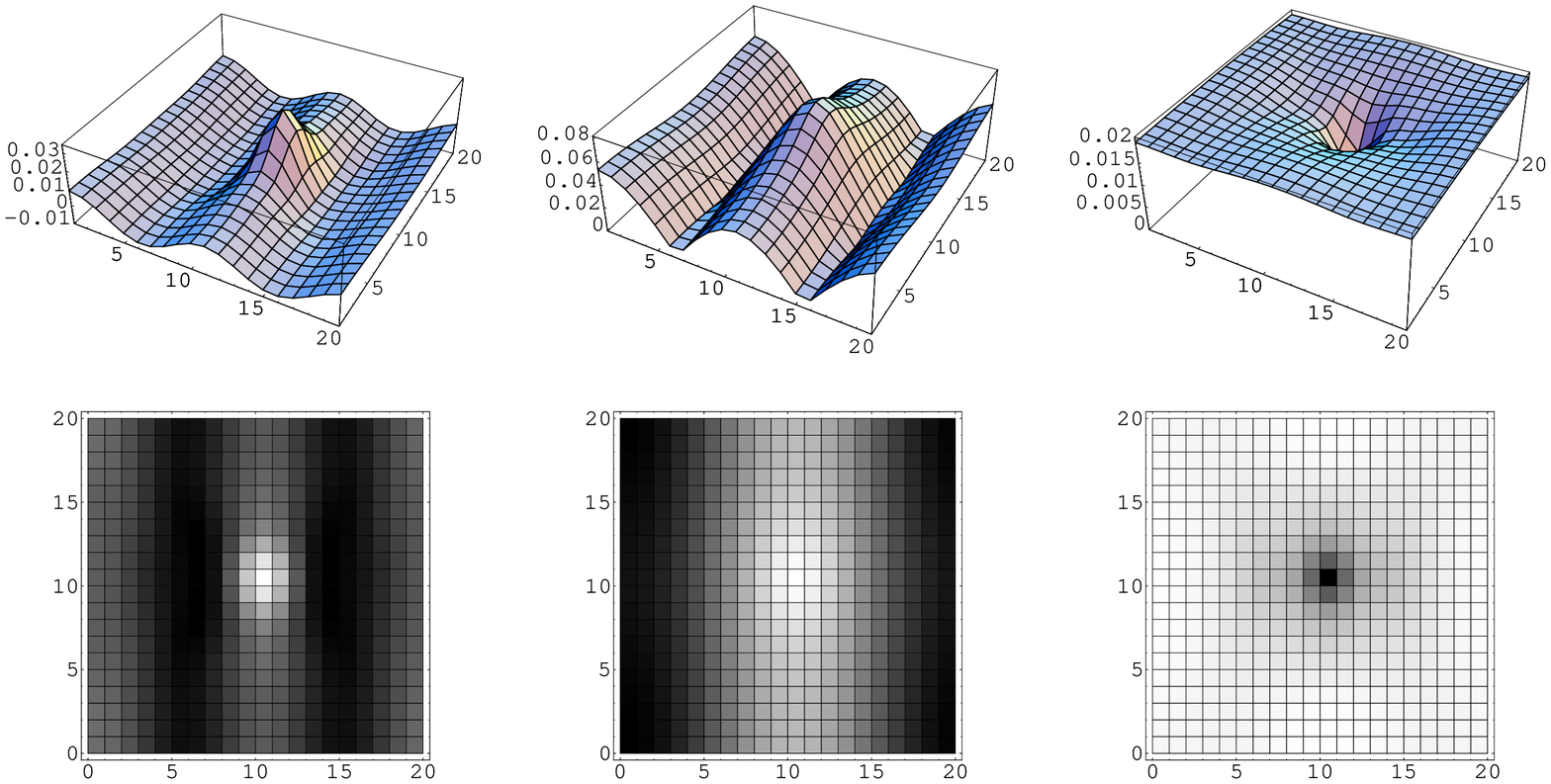}
\caption{\label{fig:stripes} From left to right, electron number density $\delta \expn{n_i} = \expn{n_i} - 0.875$, AFM order $m_i$, and $d$SC order $|\Delta_i|$ for an interaction strength $J=1.3$ and a LRC strength $V=0.35$ on one half of a $20 \times 40$ unit cell. The bottom row shows 2D density plots of the same. Both the electron density and the AFM order show a strong anisotropy, while the $d$SC order is only mildly affected.  Note that, in contrast to the $J=1.15$ results for $V=0.35$, the AFM order survives and the charge density is negative at the vortex core.}
\ewfig
the AFM and the charge density are strongly one dimensional, extending throughout the unit cell along the $y$ direction while oscillating in the $x$ direction with a period of $20 a_0$ for the AFM order and $10 a_0$ for the charge density order.  The period of this oscillation is set by the size of the unit cell.  We note that other periodicities and more complex orders are accessible within our model if terms such as a next-nearest-neighbor hopping are added.\cite{zhu02, ting02}  The $d$SC order is only slightly affected and, for the most part, retains its 2D structure.  The AFM is strongest in the vortex core and has the opposite sign at the edges of the unit cell.  The magnitude of the AFM at the edges is almost as large as it is in the center ``stripe'' and has, surprisingly, been \textit{enhanced} by the LRC interaction.  The charge density tracks the magnitude of the AFM and shows none of the rich, short-range, two-dimensional structure seen at $J=1.15$ in the presence of the LRC interaction.  With $J=1.3$, the main effect of the LRC interaction (apart from reducing the magnitude of the AFM and the charge density at the core) is to strongly enhance the one-dimensional ordering that was only weakly developed at $V=0$.  We suggest the following reasons for this phenomenology.  
First, as the exchange interaction strength $J$ is increased, the domain walls at which the AFM order parameter $m_i$ changes sign cost increasing amounts of energy.  As the AFM is strengthened, this energy cost creates a tendency to straighten out and shorten these domain walls.  We have also noticed a much weaker tendency toward anisotropy at $J=1.15$ in larger unit cells with $24 \times 48$ and $26 \times 52$ sites, that leads to strengthening of the AFM (as can be seen in Fig.\ \ref{fig:af_den_size}).  This anisotropy is not obvious at $V=0$, and is only weakly apparent for $V > 0$.  This unit-cell size dependence suggests that the correlation length $\ell_m$, which decreases with increasing $J$, may also be playing a role.  Earlier work on stripes within SO(5) theory\cite{veillette99} showed that phase-separated AFM and $d$SC regions change from a droplet phase to one with alternating stripes when the AFM and $d$SC regions occupy roughly equal areas of the superconductor.  The anisotropy observed here may be partly driven by coincidence of $\ell_m$ with the length of the unit cell, such that the AFM order parameter would like to occupy half of the unit cell area.  
Secondly, the LRC repulsion reduces the charge density at the vortex core and creates a hole-rich screening region immediately next to the core that screens the vortex charge from the rest of the unit cell.  The charge that is moved away from the core is thus pushed to the edges of the unit cell.  This balancing of the charge reduces the AFM in the center and strengthens it along the edges of the unit cell.  The LRC interaction does its best to create a uniform charge density and strongly enhances the 1D character seeded by the AFM order by washing out modulations in the charge density along the $y$ direction.  Further increase of the LRC interaction will maintain this stripelike order, but will reduce the magnitude of both the charge and AFM oscillations.

%======================================================================
\subsection{Local density of states}
\label{ssec:ldos}

The local density of states provides a basis for comparison of theoretical results with powerful experimental tools.  As is well known, the zero-bias conductance peak (ZBCP) predicted to exist in the vortex core by standard BCS theory for $d$-wave superconductors\cite{wang95} has not been experimentally observed.  Scanning-tunneling microscopy (STM) measurements of the vortex state of YBCO (\Ref{maggio95}) and of BSCCO (\Refs{renner98, pan00, hoffman02}) show a complete absence of the ZBCP in the vortex core. Low-energy structure in the conductance at $\pm 5.5$ meV in YBCO (\Ref{maggio95}) and at $\pm 7$ meV in BSCCO (\Refs{pan00, hoffman02}) is observed in the vortex core spectra, and is seen to persist outside of the core to distances much longer than the coherence length.  

We present our results for the local density of states in the vortex core region in Fig.\ \ref{fig:ldos}. 
\bfig{p}
\fig{0.48}{figure8.eps}
\caption{\label{fig:ldos} The local density of states at the center of the vortex core (top), away from the vortex core along the node direction (middle), and averaged over all sites within a coherence length from the vortex center (bottom), for $J=1.15$ and $n_{\mathrm{ave}}=0.875$ on a $26 \times 52$ unit cell. Solid lines are for $V=0.15$ and dashed lines are for $V=0$. The lines have been shifted vertically for clarity. }
\efig
 In agreement with previous results,\cite{andersen00, ogata99, ting01, ghosal02} we find that the presence of antiferromagnetic order in the vortex core provides a mechanism for splitting of the ZBCP, both in the presence and absence of the LRC interaction.   More importantly, the LRC interaction changes the spectrum at low energy, leading to excellent qualitative agreement with experimental results.  The main features in the local density of states attributable to the LRC interaction are the following: the unphysical spin gap at $E\sim 0.5~t$ is washed out for $V=0.15$ due to weakening of the AFM order; the asymmetry in height between the split zero-bias conductance peaks is reduced; and the peak splitting is found to decrease linearly with $|m_{\mathrm{core}}|$, as expected for the appearance of an order parameter that breaks time-reversal symmetry.\cite{matsumoto95_1, matsumoto95_2, sigrist95, rainer97, fogelstrom97}  The bottom trace of Fig.\ \ref{fig:ldos} shows the local density of states averaged over all sites within the vortex core.  Given the simplicity of our minimal mean-field model, the agreement at $V=0.15$ with STM results on YBCO (\Ref{maggio95}) is remarkable.  Almost all of the qualitative low-energy features of the experimental data, including the appearance of slightly asymmetric peaks at $E \sim \Delta/4$, with higher energy asymmetric ``humps'' (remnants of the strongly suppressed coherence peaks), are reproduced here.  However, the weaker low-energy ``shoulders''  observed in BSCCO (\Refs{renner98, pan00}) are not consistent with our data.  Since the coherence length is much smaller in BSCCO than in YBCO, it is more likely that mean-field theory will break down in the vortex cores of BSCCO, and that strong-correlation effects not captured in our model are responsible for the shift of weight to higher energies.  We do note, however, that the vortex-induced states at $\pm 7$ meV are seen to persist beyond the vortex core,\cite{pan00, hoffman02} in partial agreement with theory.

%=======================================================================
\section{Summary and conclusions}

We find that the negatively charged, antiferromagnetic vortex core is preserved up to a long-range Coulomb interaction strength of $V=0.35$ in the underdoped superconductor.  For $t \sim 0.5$ eV, this corresponds to a static dielectric constant for the bound charges of $\epsilon_h \sim 20$.  Experiments indicate that the static dielectric constant of the bound charges is on the order of 10 to 30+ in the undoped parent insulator\cite{samara90, chen91} and in lightly doped cuprates,\cite{varyukhin93} a range which is typical for other oxides.\cite{kastner98}  This suggests that negatively charged vortices, induced by the presence of antiferromagnetic order in the vortex core, could well be a robust feature of the cuprate superconductors in the underdoped region of the phase diagram.  It is important to note that variation in material properties, such as the chemistry and ordering of the interlayer donor regions, or the interlayer distance itself, could easily lead to non universal behavior across various families of the cuprates. In any case, one of the principal results of this paper is that \textit{static} antiferromagnetic order cannot be sustained in the absence of a negatively charged core.  The nucleation of antiferromagnetic order as the magnetic field, the doping, or the long-range Coulomb interaction strength are changed is always accompanied by a transition to a negatively charged vortex core.  On this basis, we argue that measurements of the vortex core charge could provide strong evidence for---or against---antiferromagnetic ordering in the vortex cores of {\hTc} superconductors.

Qualitative agreement of the local density of states with STM data\cite{maggio95, renner98, pan00, hoffman02} is dramatically improved when the long-range Coulomb repulsion is included in our model, suggesting that the splitting of the zero-bias conductance peak [at least in YBCO (\Ref{maggio95})] can be explained by the presence of antiferromagnetic order in the vortex core.   We find that the $\Pi$-triplet pairing amplitude is preserved outside of the vortex core, as long as there is coexistence of AFM and $d$SC orders.  We have also seen that a stripelike ordering of the antiferromagnetism and the charge density can occur when the AFM interaction strength is increased, and that this one-dimensional anisotropy is strongly enhanced by the long-range Coulomb interaction.  

At the relatively high magnetic fields to which we are limited by our approach, the vortices are close enough to each other that static AFM order, which has a correlation length $\ell \gg \xi$, is extended throughout the bulk of the sample and should be robust against the effects of fluctuations; AFM is still noticeable at the boundary of a unit cell size of $26 \times 52$ sites, which corresponds to a field of 19 T.  At much lower magnetic fields, where the intervortex distance $\ell_v \gg \ell$, fluctuations may well destroy the long range AFM order.  However, even in the absence of long-range order, it is possible that signatures of localized antiferromagnetism would be detectable at low frequencies in the vicinity of the vortex cores.

To our knowledge, only one attempt at a direct measurement of the vortex core charge in {\hTc} cuprate superconductors has been reported.\cite{kumagai01}  Nuclear quadrupole resonance (NQR) measurements of a field-induced shift in the NQR frequency, which is sensitive to the local charge density, of the copper nuclei suggest that slightly overdoped {\ybco{7}} has negatively charged vortices and underdoped {\ybcoeight} has positively charged vortices.\cite{kumagai01} The results of this experiment suggest a doping dependence of the vortex core charge that is opposite to that reported in this paper, and in seeming contradiction with the many observations of field-induced antiferromagnetic order\cite{mitrovic03, kakuyanagi03, miller02, lake01, hoffman02, lake02, katano00, khaykovich02} in the cuprates and with the doping dependence of the Hall sign.\cite{nagaoka98}  It would be interesting to see the results of these and other\cite{clayhold03} measurements of the vortex-core charge in the more highly-underdoped cuprates.  The results of this paper suggest that such measurements could well be definitive.

%==========================================================================
\begin{acknowledgments}
We would like to thank A.\ John Berlinsky for many useful discussions.  We acknowledge the financial support and use of computing facilities provided by SHARCNET (\url{http://www.sharcnet.ca}) at McMaster University.  This research was supported by the Natural Sciences and Engineering Research Council (Canada), by the Canadian Institute for Advanced Research and by the Canada Research Chairs program.  
\end{acknowledgments}

%==========================================================================
\appendix

\section{Convergence Method} \label{sec:convergence}

The long-range Coulomb interaction severely destabilizes self-consistent calculation of the order parameters, by causing divergent long-wavelength oscillations in the charge density from one iteration to the next. In order to suppress these divergent oscillations we tried a number of convergence methods, in increasing order of sophistication, which we review below.

The simplest way to suppress these oscillations is to linearly mix the input ($\ket{\chi_\inrm^{(m)}}$) and the output ($\ket{\chi_\outrm^{(m)}}$) from the current iteration to generate the input for the next iteration $\ket{\chi_{\inrm}^{(m+1)}} = (1-\alpha) \ket{\chi_{\inrm}^{(m)}} + \alpha \ket{\chi_{\outrm}^{(m)}}$.  While this method has been used with some success in other self-consistent calculations (see \Ref{pickett}, and references therein) it is unable to suppress the divergent oscillations in the charge density in the BdG calculations of this paper for any reasonable values of $\alpha$ (the reasons for this failure are discussed in \Ref{raczkowski01}).  

A more sophisticated approach is Broyden's method.\cite{pickett}  Let
\begin{equation}
\ket{F^{(m)}} = \ket{\chi^{(m)}_{\outrm}} -
\ket{\chi^{(m)}_{\inrm}},
\end{equation}
where $\ket{F} = \ket{0}$ at convergence.  Expanding $\ket{F}$ to linear order near convergence gives
\begin{equation} \label{eq:linear}
\ket{F} \approx
\ket{F^{(m)}}-\mathbb{J}^{(m)}(\ket{\chi_{\inrm}}-\ket{\chi_{\inrm}^{(m)}}),
\end{equation}
where the Jacobian
\begin{equation} \label{eq:jac_def}
\mathbb{J}_{i,j}^{(m)} = -\frac{\delta\ket{F^{(m)}}_i}
{\delta\ket{\chi^{(m)}_{\inrm}}_j}.
\end{equation}
If convergence occurs at iteration $m+1$, such that $\ket{F^{(m+1)}}=0$, then
\begin{equation} \label{eq:update}
\ket{\chi_{\inrm}^{(m+1)}} = \ket{\chi_{\inrm}^{(m)}} +
\mathbb{G}^{(m)}\ket{F^{(m)}},
\end{equation}
where $\mathbb{G}=[\mathbb{J}]^{-1}$ is the inverse Jacobian. This equation is used to generate a new set of inputs after each iteration.  The goal of Broyden's method is to improve the inverse Jacobian at each iteration so that the new inputs $\ket{\chi_{\inrm}^{(m+1)}}$ are as close to convergence as possible.

Define, for convenience and for numerical accuracy,
\begin{gather}
\ket{\Delta F^{(m)}} = \frac{\ket{F^{(m)}} - \ket{F^{(m-1)}}}
{|~\ket{\chi^{(m)}_{\inrm}}-\ket{\chi^{(m-1)}_{\inrm}}~|},
\\
\ket{\Delta \chi^{(m)}} = \frac{\ket{\chi^{(m)}_{\inrm}} -
\ket{\chi^{(m-1)}_{\inrm}}}
{|~\ket{\chi^{(m)}_{\inrm}}-\ket{\chi^{(m-1)}_{\inrm}}~|},
\end{gather}
such that $\braket{\Delta \chi^{(m)}}=1$.  At convergence $\mathbb{G}^{(m)} = \mathbb{G}^{(m-1)}$.  This suggests that the path to convergence can be found by minimizing the change
\begin{equation} \label{eq:errorJ}
E = ||\mathbb{G}^{(m)} - \mathbb{G}^{(m-1)}||
\end{equation}
 (where $||...||$ denotes the Frobenius norm) in the inverse Jacobian from one iteration to the  next, subject to the constraint [from \Eq{eq:linear}] that
\begin{equation}
\ket{\Delta \chi^{(m)}} = - \mathbb{G}^{(m)} \ket{\Delta F^{(m)}}.
\end{equation}
This minimization leads to a recursion relation for the inverse Jacobian ($\mathbb{G}^{(m+1)}$) that is the basis of Broyden's method.\cite{pickett}  While the use of Broyden's method does go some way towards suppressing oscillations in the charge density from iteration to iteration, we found that a long-wavelength oscillation (with a period in ``time'' of two iterations) in the charge density would slowly build up over many iterations and cause the system to diverge.

A higher order approach is the modified Broyden's method, first introduced by Vanderbilt and Louie (VL).\cite{vanderbilt84}  The problem with Broyden's method is that the inverse Jacobian $\mathbb{G}^{(m)}$ has not been required to satisfy
\begin{equation} \label{eq:constraint}
\ket{\Delta \chi^{(n)}} = -\mathbb{G}^{(m)} \ket{\Delta F^{(n)}}
\end{equation}
for all previous iterations $n < m$, as it should.  

Following VL.\cite{vanderbilt84} information from all previous iterations can be introduced by doing a least-squares minimization of the following ``error function:''
\begin{equation} \label{eq:error}
E = \omega_0^2||\mathbb{J}^{(m)}-\mathbb{J}^{(0)}|| +
    \sum_{n=1}^m \omega_n^2 |~ \ket {\Delta F^{(n)}} + \mathbb{J}^{(m)} \ket {\Delta \chi^{(n)}} ~|^2,
\end{equation}
where $\omega_n$ represents the weight given to the information from the $n$th iteration.  As an aside, we note that we tried various weighting schemes and found that choosing 
\begin{equation}
\omega_n = - \log \left(\frac{\braket{F^{(m)}}}{\braket{\chi_{\inrm}^{(m)}}} \right),~\omega_0 = 0.01,
\end{equation}
which gives more weight to the information from iterations in which the difference $\ket{\chi_{\outrm}} - \ket{\chi_{\inrm}}$ is small, works well for our particular problem.  The new error function introduced by VL allows the minimization of both the change and the error [based on {\Eq{eq:constraint}}] in the Jacobian from iteration to iteration. Note that we are now considering the change in the Jacobian and not the inverse Jacobian, and that the change in the Jacobian $||\mathbb{J}^{(m)}-\mathbb{J}^{(0)}||$, is now defined relative to $\mathbb{J}^{(0)}$. This may seem like a step backwards, since it is the inverse Jacobian that is needed to generate the inputs for the next iteration [as in \Eq{eq:update}].  This point was made by Johnson,\cite{johnson} who introduced a numerically more efficient version of the modified Broyden's method by defining an error function in terms of the inverse Jacobian $\mathbb{G}$.  However, we have found that, for our particular problem, using the error function based on the Jacobian leads to a procedure which requires far fewer iterations to reach self-consistency.  Furthermore, by making the standard assumption\cite{vanderbilt84, johnson} that the initial Jacobian $\mathbb{J}^{(0)}=\mathbb{I}/\alpha$ (meaning that linear mixing is used to construct the first guess) we are able to use the numerical innovations of Johnson to greatly enhance the efficiency of the method of VL by avoiding any explicit calculations of the inverse.

We start by minimizing the error function [\Eq{eq:error}] with respect to the new Jacobian.  Setting $\partial E/ \partial \mathbb{J}_{ij}^{(m)}=0$ gives (as in \Ref{vanderbilt84})
\begin{equation} \label{eq:VLjacobian}
\mathbb{J}^{(m)}= \Gamma^{(m)} [\beta^{(m)}]^{-1},
\end{equation}
where
\begin{equation}
\Gamma^{(m)}=\mathbb{J}^{(0)} - \sum_{n=1}^m \frac{\omega_n^2}{\omega_0^2} \ket{\Delta F^{(n)}}\bra{\Delta \chi^{(n)}}
\end{equation}
and
\begin{equation} \label{eq:beta}
\beta^{(m)}= \mathbb{I} + \sum_{n=1}^m \frac{\omega_n^2}{\omega_0^2} \ketbra{\Delta \chi^{(n)}}.
\end{equation}
In order to update the inputs for the next iteration, as in \Eq{eq:update}, one needs the inverse Jacobian $\mathbb{G}^{(m)} = \beta^{(m)}[\Gamma^{(m)}]^{-1}$.  Assuming that $\mathbb{J}^{(0)}=\mathbb{I}/\alpha$, 
\begin{equation}
[\Gamma^{(m)}]^{-1} = \alpha \left[ \mathbb{I}-\sum_{n=1}^m \alpha \frac{\omega_n^2}{\omega_0^2} \ket{\Delta F^{(n)}} \bra{\Delta \chi^{(n)}} \right]^{-1}.
\end{equation}
Borrowing an idea from Johnson,\cite{johnson} we expand $[\Gamma^{(m)}]^{-1}$ to infinite order in the vectors $\ket{\Delta F^{(n)}}$ and $\ket{\Delta \chi^{(n)}}$ and then resum: 
\begin{equation} \label{eq:gamma}
[\Gamma^{(m)}]^{-1}=\alpha \mathbb{I} + \alpha \sum_{n,\ell=1}^m \omega_n \omega_\ell \gamma_{n\ell} \ket{\Delta F^{(n)}} \bra{\Delta \chi^{(\ell)}},
\end{equation}
where $\gamma_{n\ell} = [\tfrac{\omega_0^2}{\alpha}\mathbb{I}-a]^{-1}_{n\ell}$ and $a_{ij}=\omega_i \omega_j \braketII{\Delta \chi^{(i)}}{\Delta F^{(j)}}$.  This procedure exchanges the inversion of a large $N \times N$ matrix for that of a much smaller $m \times m$ matrix. The inverse Jacobian is thus
\begin{equation} \label{eq:finalG}
\mathbb{G}^{(m)} = \mathbb{G}^{(0)} + \sum_{k,\ell=1}^m \omega_\ell \gamma_{k\ell} \ketbraII{u^{(k)}}{\Delta \chi^{(\ell)}}, 
\end{equation}
where $\ket{u^{(k)}}= \omega_k (\mathbb{G}^{(0)} \ket{\Delta F^{(k)}} + \ket{\Delta \chi^{(k)}})$ and where we have used the identity
\begin{equation}
a_{kn} \gamma_{n\ell} = -\delta_{k\ell} + \frac{\omega_0^2}{\alpha}\gamma_{k\ell}
\end{equation}
to simplify the product $\beta^{(m)}[\Gamma^{(m)}]^{-1}$. 

Substituting \Eq{eq:finalG} into the update (\ref{eq:update}) gives
\begin{equation}
\ket{\chi^{(m+1)}_{\inrm}}=\ket{\chi^{(m)}_{\inrm}}+\alpha \ket{F^{(m)}} + \sum_{k=1}^m d^{(m)}_k \ket{u^{(k)}},
\end{equation}
where 
\begin{equation}
d^{(m)}_k = \sum_{\ell =1}^m \gamma_{k\ell}~ c^{(\ell)}
\end{equation}
and
\begin{equation}
c^{(\ell)} = \omega_{\ell}\braketII{\Delta \chi^{(\ell)}}{F^{(m)}}
\end{equation}
have been introduced for numerical convenience. The modified Broyden's method described here should be generally applicable to problems other than the one discussed in this paper.
\footnote{We have posted a copy of the Fortran77 subroutine used to implement the modified Broyden's method described here online at \texttt{www.danielk.ca/code.html}.}

The use of the modified Broyden's method eliminates the divergent oscillations in the charge density by taking information from all previous iterations to construct the input for the next iteration
\cite{atkinson04}${}^,$
\footnote{At the time of writing we became aware of a recently-developed convergence-method, specific to the self-consistent calculation of Coulomb potentials and known as the Pulay-Thomas-Fermi mixing scheme\cite{raczkowski01}, which has been used in a study of dopant-induced inhomogeneity in the cuprates\cite{atkinson04}.}.  
Since we start with a converged solution for $V = 0$ and since the largest changes at $V > 0$ are to the charge density we restrict application of the modified Broyden's method to updates of the Hartree shift.  The inputs for all other order parameters are taken to be the outputs of the current iteration.  We found that it was necessary to fix the antiferromagnetic order parameter---which strongly influences the charge density---and allow the Hartree shift to partially converge over several iterations using the modified Broyden's method.  Temporarily fixing the antiferromagnetism keeps the convergent endpoint of the Hartree shift within reach of the modified Broyden's method (which is based on the assumption that the starting state is not too far from convergence).  Once the Hartree shift has partially converged, we allow the antiferromagnetism to change and then \textit{restart} the modified Broyden's method for the Hartree shift by setting $\mathbb{J} = \mathbb{I}/\alpha$ and taking as initial inputs the most recent outputs. In this way the system proceeds stepwise towards a convergent endpoint without exceeding the reach of the modified Broyden's method in any given step. 

%=====================================================================


\begin{thebibliography}{76}

\expandafter\ifx\csname natexlab\endcsname\relax\def\natexlab#1{#1}\fi
\expandafter\ifx\csname bibnamefont\endcsname\relax
  \def\bibnamefont#1{#1}\fi
\expandafter\ifx\csname bibfnamefont\endcsname\relax
  \def\bibfnamefont#1{#1}\fi
\expandafter\ifx\csname citenamefont\endcsname\relax
  \def\citenamefont#1{#1}\fi
\expandafter\ifx\csname url\endcsname\relax
  \def\url#1{\texttt{#1}}\fi
\expandafter\ifx\csname urlprefix\endcsname\relax\def\urlprefix{URL }\fi
\providecommand{\bibinfo}[2]{#2}
\providecommand{\eprint}[2][]{\url{#2}}

\bibitem[{\citenamefont{Maggio-Aprile et~al.}(1995)\citenamefont{Maggio-Aprile,
  Renner, Erb, Walker, and Fischer}}]{maggio95}
\bibinfo{author}{\bibfnamefont{I.}~\bibnamefont{Maggio-Aprile}},
  \bibinfo{author}{\bibfnamefont{C.}~\bibnamefont{Renner}},
  \bibinfo{author}{\bibfnamefont{A.}~\bibnamefont{Erb}},
  \bibinfo{author}{\bibfnamefont{E.}~\bibnamefont{Walker}}, \bibnamefont{and}
  \bibinfo{author}{\bibfnamefont{{\O}.}~\bibnamefont{Fischer}},
  \bibinfo{journal}{Phys.\ Rev.\ Lett.} \textbf{\bibinfo{volume}{75}},
  \bibinfo{pages}{2754} (\bibinfo{year}{1995}).

\bibitem[{\citenamefont{Renner et~al.}(1998)\citenamefont{Renner, Revaz,
  Kadowaki, Maggio-Aprile, and Fischer}}]{renner98}
\bibinfo{author}{\bibfnamefont{C.}~\bibnamefont{Renner}},
  \bibinfo{author}{\bibfnamefont{B.}~\bibnamefont{Revaz}},
  \bibinfo{author}{\bibfnamefont{K.}~\bibnamefont{Kadowaki}},
  \bibinfo{author}{\bibfnamefont{I.}~\bibnamefont{Maggio-Aprile}},
  \bibnamefont{and}
  \bibinfo{author}{\bibfnamefont{{\O}.}~\bibnamefont{Fischer}},
  \bibinfo{journal}{Phys.\ Rev.\ Lett.} \textbf{\bibinfo{volume}{80}},
  \bibinfo{pages}{3606} (\bibinfo{year}{1998}).

\bibitem[{\citenamefont{Pan et~al.}(2000)\citenamefont{Pan, Hudson, Gupta, Ng,
  Eisaki, Uchida, and Davis}}]{pan00}
\bibinfo{author}{\bibfnamefont{S.~H.} \bibnamefont{Pan}},
  \bibinfo{author}{\bibfnamefont{E.~W.} \bibnamefont{Hudson}},
  \bibinfo{author}{\bibfnamefont{A.~K.} \bibnamefont{Gupta}},
  \bibinfo{author}{\bibfnamefont{K.-W.} \bibnamefont{Ng}},
  \bibinfo{author}{\bibfnamefont{H.}~\bibnamefont{Eisaki}},
  \bibinfo{author}{\bibfnamefont{S.}~\bibnamefont{Uchida}}, \bibnamefont{and}
  \bibinfo{author}{\bibfnamefont{J.~C.} \bibnamefont{Davis}},
  \bibinfo{journal}{Phys.\ Rev.\ Lett.} \textbf{\bibinfo{volume}{85}},
  \bibinfo{pages}{1536} (\bibinfo{year}{2000}).

\bibitem[{\citenamefont{Hoffman et~al.}(2002)\citenamefont{Hoffman, Hudson,
  Lang, Madhavan, Eisaki, Uchida, and Davis}}]{hoffman02}
\bibinfo{author}{\bibfnamefont{J.~E.} \bibnamefont{Hoffman}},
  \bibinfo{author}{\bibfnamefont{E.~W.} \bibnamefont{Hudson}},
  \bibinfo{author}{\bibfnamefont{K.~M.} \bibnamefont{Lang}},
  \bibinfo{author}{\bibfnamefont{V.}~\bibnamefont{Madhavan}},
  \bibinfo{author}{\bibfnamefont{H.}~\bibnamefont{Eisaki}},
  \bibinfo{author}{\bibfnamefont{S.}~\bibnamefont{Uchida}}, \bibnamefont{and}
  \bibinfo{author}{\bibfnamefont{J.~C.} \bibnamefont{Davis}},
  \bibinfo{journal}{Science} \textbf{\bibinfo{volume}{295}},
  \bibinfo{pages}{466} (\bibinfo{year}{2002}).

\bibitem[{\citenamefont{Curro et~al.}(2000)\citenamefont{Curro, Milling, Haase,
  and Slichter}}]{curro00}
\bibinfo{author}{\bibfnamefont{N.~J.} \bibnamefont{Curro}},
  \bibinfo{author}{\bibfnamefont{C.}~\bibnamefont{Milling}},
  \bibinfo{author}{\bibfnamefont{J.}~\bibnamefont{Haase}}, \bibnamefont{and}
  \bibinfo{author}{\bibfnamefont{C.~P.} \bibnamefont{Slichter}},
  \bibinfo{journal}{Phys.\ Rev.\ B} \textbf{\bibinfo{volume}{62}},
  \bibinfo{pages}{3473} (\bibinfo{year}{2000}).

\bibitem[{\citenamefont{Mitrovi{\'c} et~al.}(2001)\citenamefont{Mitrovi{\'c},
  Sigmund, Eschrig, Bachman, Halperin, Reyes, Kuhns, and Moulton}}]{mitrovic01}
\bibinfo{author}{\bibfnamefont{V.~F.} \bibnamefont{Mitrovi{\'c}}},
  \bibinfo{author}{\bibfnamefont{E.~E.} \bibnamefont{Sigmund}},
  \bibinfo{author}{\bibfnamefont{M.}~\bibnamefont{Eschrig}},
  \bibinfo{author}{\bibfnamefont{H.~N.} \bibnamefont{Bachman}},
  \bibinfo{author}{\bibfnamefont{W.~P.} \bibnamefont{Halperin}},
  \bibinfo{author}{\bibfnamefont{A.~P.} \bibnamefont{Reyes}},
  \bibinfo{author}{\bibfnamefont{P.}~\bibnamefont{Kuhns}}, \bibnamefont{and}
  \bibinfo{author}{\bibfnamefont{W.~G.} \bibnamefont{Moulton}},
  \bibinfo{journal}{Nature} \textbf{\bibinfo{volume}{413}},
  \bibinfo{pages}{501} (\bibinfo{year}{2001}).

\bibitem[{\citenamefont{Kakuyanagi et~al.}(2002)\citenamefont{Kakuyanagi, Kumagai, and Matsuda}}]{kakuyanagi02}
\bibinfo{author}{\bibfnamefont{K.}~\bibnamefont{Kakuyanagi}},
  \bibinfo{author}{\bibfnamefont{K.~I.}~\bibnamefont{Kumagai}},
  \bibnamefont{and} \bibinfo{author}{\bibfnamefont{Y.}~\bibnamefont{Matsuda}},
  \bibinfo{journal}{Phys.\ Rev.\ B} \textbf{\bibinfo{volume}{65}},
  \bibinfo{pages}{060503(R)} (\bibinfo{year}{2002}).

\bibitem[{\citenamefont{Mitrovi{\'c} et~al.}(2003)\citenamefont{Mitrovi{\'c},
  Sigmund, Halperin, Reyes, Kuhns, and Moulton}}]{mitrovic03}
\bibinfo{author}{\bibfnamefont{V.~F.} \bibnamefont{Mitrovi{\'c}}},
  \bibinfo{author}{\bibfnamefont{E.~E.} \bibnamefont{Sigmund}},
  \bibinfo{author}{\bibfnamefont{W.~P.} \bibnamefont{Halperin}},
  \bibinfo{author}{\bibfnamefont{A.~P.} \bibnamefont{Reyes}},
  \bibinfo{author}{\bibfnamefont{P.}~\bibnamefont{Kuhns}}, \bibnamefont{and}
  \bibinfo{author}{\bibfnamefont{W.~G.} \bibnamefont{Moulton}},
  \bibinfo{journal}{Phys.\ Rev.\ B} \textbf{\bibinfo{volume}{67}},
  \bibinfo{pages}{220503(R)} (\bibinfo{year}{2003}).

\bibitem[{\citenamefont{Kakuyanagi et~al.}(2003)\citenamefont{Kakuyanagi,
  Kumagai, Matsuda, and Hasegawa}}]{kakuyanagi03}
\bibinfo{author}{\bibfnamefont{K.}~\bibnamefont{Kakuyanagi}},
  \bibinfo{author}{\bibfnamefont{K.}~\bibnamefont{Kumagai}},
  \bibinfo{author}{\bibfnamefont{Y.}~\bibnamefont{Matsuda}}, \bibnamefont{and}
  \bibinfo{author}{\bibfnamefont{M.}~\bibnamefont{Hasegawa}},
  \bibinfo{journal}{Phys.\ Rev.\ Lett.} \textbf{\bibinfo{volume}{90}},
  \bibinfo{pages}{197003} (\bibinfo{year}{2003}).

\bibitem[{\citenamefont{Vershinin et~al.}(2004)\citenamefont{Vershinin, Misra,
  Ono, Abe, Ando, and Yazdani}}]{vershinin04}
\bibinfo{author}{\bibfnamefont{M.}~\bibnamefont{Vershinin}},
  \bibinfo{author}{\bibfnamefont{S.}~\bibnamefont{Misra}},
  \bibinfo{author}{\bibfnamefont{S.}~\bibnamefont{Ono}},
  \bibinfo{author}{\bibfnamefont{Y.}~\bibnamefont{Abe}},
  \bibinfo{author}{\bibfnamefont{Y.}~\bibnamefont{Ando}}, \bibnamefont{and}
  \bibinfo{author}{\bibfnamefont{A.}~\bibnamefont{Yazdani}},
  \bibinfo{journal}{Science} \textbf{\bibinfo{volume}{303}},
  \bibinfo{pages}{1995} (\bibinfo{year}{2004}).

\bibitem[{\citenamefont{Howald et~al.}(2003)\citenamefont{Howald, Eisaki,
  Kaneko, Greven, and Kapitulnik}}]{howald03}
\bibinfo{author}{\bibfnamefont{C.}~\bibnamefont{Howald}},
  \bibinfo{author}{\bibfnamefont{H.}~\bibnamefont{Eisaki}},
  \bibinfo{author}{\bibfnamefont{N.}~\bibnamefont{Kaneko}},
  \bibinfo{author}{\bibfnamefont{M.}~\bibnamefont{Greven}}, \bibnamefont{and}
  \bibinfo{author}{\bibfnamefont{A.}~\bibnamefont{Kapitulnik}},
  \bibinfo{journal}{Phys.\ Rev.\ B} \textbf{\bibinfo{volume}{67}},
  \bibinfo{pages}{014533} (\bibinfo{year}{2003}).

\bibitem[{\citenamefont{McElroy et~al.}(2004)\citenamefont{McElroy, Lee,
  Hoffman, Lang, Hudson, Eisaki, Uchida, Lee, and Davis}}]{davis04}
\bibinfo{author}{\bibfnamefont{K.}~\bibnamefont{McElroy}},
  \bibinfo{author}{\bibfnamefont{D.-H.} \bibnamefont{Lee}},
  \bibinfo{author}{\bibfnamefont{J.~E.} \bibnamefont{Hoffman}},
  \bibinfo{author}{\bibfnamefont{K.~M.} \bibnamefont{Lang}},
  \bibinfo{author}{\bibfnamefont{E.~W.} \bibnamefont{Hudson}},
  \bibinfo{author}{\bibfnamefont{E.}~\bibnamefont{Eisaki}},
  \bibinfo{author}{\bibfnamefont{S.}~\bibnamefont{Uchida}},
  \bibinfo{author}{\bibfnamefont{J.}~\bibnamefont{Lee}}, \bibnamefont{and}
  \bibinfo{author}{\bibfnamefont{J.~C.} \bibnamefont{Davis}}
  (\bibinfo{year}{2004}), \eprint{arXiv:cond-mat/0404005}.

\bibitem[{\citenamefont{Hanaguri et~al.}(2004)\citenamefont{Hanaguri, Lupien,
  Kohsaka, Lee, Azuma, Takano, Takagi, and Davis}}]{hanaguri04}
\bibinfo{author}{\bibfnamefont{T.}~\bibnamefont{Hanaguri}},
  \bibinfo{author}{\bibfnamefont{C.}~\bibnamefont{Lupien}},
  \bibinfo{author}{\bibfnamefont{Y.}~\bibnamefont{Kohsaka}},
  \bibinfo{author}{\bibfnamefont{D.-H.} \bibnamefont{Lee}},
  \bibinfo{author}{\bibfnamefont{M.}~\bibnamefont{Azuma}},
  \bibinfo{author}{\bibfnamefont{M.}~\bibnamefont{Takano}},
  \bibinfo{author}{\bibfnamefont{H.}~\bibnamefont{Takagi}}, \bibnamefont{and}
  \bibinfo{author}{\bibfnamefont{J.~C.} \bibnamefont{Davis}},
  \bibinfo{journal}{Nature} \textbf{\bibinfo{volume}{430}},
  \bibinfo{pages}{1001} (\bibinfo{year}{2004}).

\bibitem[{\citenamefont{Chen et~al.}(2002{\natexlab{a}})\citenamefont{Chen, Hu,
  Capponi, Arrigoni, and Zhang}}]{zhang02}
\bibinfo{author}{\bibfnamefont{H.-D.} \bibnamefont{Chen}},
  \bibinfo{author}{\bibfnamefont{J.-P.} \bibnamefont{Hu}},
  \bibinfo{author}{\bibfnamefont{S.}~\bibnamefont{Capponi}},
  \bibinfo{author}{\bibfnamefont{E.}~\bibnamefont{Arrigoni}}, \bibnamefont{and}
  \bibinfo{author}{\bibfnamefont{S.-C.} \bibnamefont{Zhang}},
  \bibinfo{journal}{Phys.\ Rev.\ Lett.} \textbf{\bibinfo{volume}{89}},
  \bibinfo{pages}{137004} (\bibinfo{year}{2002}{\natexlab{a}}).

\bibitem[{\citenamefont{Chen et~al.}(2004)\citenamefont{Chen, Vafek, Yazdani,
  and Zhang}}]{zhang04}
\bibinfo{author}{\bibfnamefont{H.-D.} \bibnamefont{Chen}},
  \bibinfo{author}{\bibfnamefont{O.}~\bibnamefont{Vafek}},
  \bibinfo{author}{\bibfnamefont{A.}~\bibnamefont{Yazdani}}, \bibnamefont{and}
  \bibinfo{author}{\bibfnamefont{S.-C.} \bibnamefont{Zhang}}
  (\bibinfo{year}{2004}), \eprint{arXiv:cond-mat/0402323}.

\bibitem[{\citenamefont{Te\u{s}anovi\'{c}}(2004)}]{tesanovic04}
\bibinfo{author}{\bibfnamefont{Z.}~\bibnamefont{Te\u{s}anovi\'{c}}}
  (\bibinfo{year}{2004}), \eprint{arXiv:cond-mat/0405235}.

\bibitem[{\citenamefont{Kivelson et~al.}(2003)\citenamefont{Kivelson, Bindloss,
  Fradkin, Oganesyan, Tranquada, Kapitulnik, and Howald}}]{kivelson03}
\bibinfo{author}{\bibfnamefont{S.~A.} \bibnamefont{Kivelson}},
  \bibinfo{author}{\bibfnamefont{I.~P.} \bibnamefont{Bindloss}},
  \bibinfo{author}{\bibfnamefont{E.}~\bibnamefont{Fradkin}},
  \bibinfo{author}{\bibfnamefont{V.}~\bibnamefont{Oganesyan}},
  \bibinfo{author}{\bibfnamefont{J.~M.} \bibnamefont{Tranquada}},
  \bibinfo{author}{\bibfnamefont{A.}~\bibnamefont{Kapitulnik}},
  \bibnamefont{and} \bibinfo{author}{\bibfnamefont{C.}~\bibnamefont{Howald}},
  \bibinfo{journal}{Rev.\ Mod.\ Phys.} \textbf{\bibinfo{volume}{75}},
  \bibinfo{pages}{1201} (\bibinfo{year}{2003}).

\bibitem[{\citenamefont{Norman}(2004)}]{norman04}
\bibinfo{author}{\bibfnamefont{M.}~\bibnamefont{Norman}},
  \bibinfo{journal}{Science} \textbf{\bibinfo{volume}{303}},
  \bibinfo{pages}{1985} (\bibinfo{year}{2004}).

\bibitem[{\citenamefont{Polkovnikov et~al.}(2002)\citenamefont{Polkovnikov,
  Vojta, and Sachdev}}]{sachdev02_rapid}
\bibinfo{author}{\bibfnamefont{A.}~\bibnamefont{Polkovnikov}},
  \bibinfo{author}{\bibfnamefont{M.}~\bibnamefont{Vojta}}, \bibnamefont{and}
  \bibinfo{author}{\bibfnamefont{S.}~\bibnamefont{Sachdev}},
  \bibinfo{journal}{Phys.\ Rev.\ B} \textbf{\bibinfo{volume}{65}},
  \bibinfo{pages}{220509(R)} (\bibinfo{year}{2002}).

\bibitem[{\citenamefont{Fu et~al.}(2004)\citenamefont{Fu, Davis, and
  Lee}}]{fu04}
\bibinfo{author}{\bibfnamefont{H.~C.} \bibnamefont{Fu}},
  \bibinfo{author}{\bibfnamefont{J.~C.} \bibnamefont{Davis}}, \bibnamefont{and}
  \bibinfo{author}{\bibfnamefont{D.-H.} \bibnamefont{Lee}}
  (\bibinfo{year}{2004}), \eprint{arXiv:cond-mat/0403001}.

\bibitem[{\citenamefont{Fong et~al.}(1999)\citenamefont{Fong, Bourges, Sidis,
  Regnault, Ivanov, Gu, Koshizuka, and Keimer}}]{keimer99}
\bibinfo{author}{\bibfnamefont{H.~F.} \bibnamefont{Fong}},
  \bibinfo{author}{\bibfnamefont{P.}~\bibnamefont{Bourges}},
  \bibinfo{author}{\bibfnamefont{Y.}~\bibnamefont{Sidis}},
  \bibinfo{author}{\bibfnamefont{L.~P.} \bibnamefont{Regnault}},
  \bibinfo{author}{\bibfnamefont{A.}~\bibnamefont{Ivanov}},
  \bibinfo{author}{\bibfnamefont{G.~D.} \bibnamefont{Gu}},
  \bibinfo{author}{\bibfnamefont{N.}~\bibnamefont{Koshizuka}},
  \bibnamefont{and} \bibinfo{author}{\bibfnamefont{B.}~\bibnamefont{Keimer}},
  \bibinfo{journal}{Nature} \textbf{\bibinfo{volume}{398}},
  \bibinfo{pages}{588} (\bibinfo{year}{1999}).

\bibitem[{\citenamefont{He et~al.}(2002)\citenamefont{He, Bourges, Sidis,
  Ulrich, Regnault, Pailh\`{e}s, Berzigiarova, Kolesnikov, and
  Keimer}}]{keimer02}
\bibinfo{author}{\bibfnamefont{H.}~\bibnamefont{He}},
  \bibinfo{author}{\bibfnamefont{P.}~\bibnamefont{Bourges}},
  \bibinfo{author}{\bibfnamefont{Y.}~\bibnamefont{Sidis}},
  \bibinfo{author}{\bibfnamefont{C.}~\bibnamefont{Ulrich}},
  \bibinfo{author}{\bibfnamefont{P.~P.} \bibnamefont{Regnault}},
  \bibinfo{author}{\bibfnamefont{S.}~\bibnamefont{Pailh\`{e}s}},
  \bibinfo{author}{\bibfnamefont{N.~S.} \bibnamefont{Berzigiarova}},
  \bibinfo{author}{\bibfnamefont{N.~N.} \bibnamefont{Kolesnikov}},
  \bibnamefont{and} \bibinfo{author}{\bibfnamefont{B.}~\bibnamefont{Keimer}},
  \bibinfo{journal}{Science} \textbf{\bibinfo{volume}{295}},
  \bibinfo{pages}{1045} (\bibinfo{year}{2002}).

\bibitem[{\citenamefont{Hinkov et~al.}(2004)\citenamefont{Hinkov, Paillh\`{e}s,
  Bourges, Sidis, Ivanov, Kulakov, Lin, Chen, Bernhard, and
  Keimer}}]{keimer04_nature}
\bibinfo{author}{\bibfnamefont{V.}~\bibnamefont{Hinkov}},
  \bibinfo{author}{\bibfnamefont{S.}~\bibnamefont{Paillh\`{e}s}},
  \bibinfo{author}{\bibfnamefont{P.}~\bibnamefont{Bourges}},
  \bibinfo{author}{\bibfnamefont{Y.}~\bibnamefont{Sidis}},
  \bibinfo{author}{\bibfnamefont{A.}~\bibnamefont{Ivanov}},
  \bibinfo{author}{\bibfnamefont{A.}~\bibnamefont{Kulakov}},
  \bibinfo{author}{\bibfnamefont{C.~T.} \bibnamefont{Lin}},
  \bibinfo{author}{\bibfnamefont{D.~P.} \bibnamefont{Chen}},
  \bibinfo{author}{\bibfnamefont{C.}~\bibnamefont{Bernhard}}, \bibnamefont{and}
  \bibinfo{author}{\bibfnamefont{B.}~\bibnamefont{Keimer}},
  \bibinfo{journal}{Nature} \textbf{\bibinfo{volume}{430}},
  \bibinfo{pages}{650} (\bibinfo{year}{2004}).

\bibitem[{\citenamefont{Pailh\`{e}s et~al.}(2004)\citenamefont{Pailh\`{e}s,
  Sidis, Bourges, Hinkov, Ivanov, Ulrich, Regnault, and Keimer}}]{keimer04}
\bibinfo{author}{\bibfnamefont{S.}~\bibnamefont{Pailh\`{e}s}},
  \bibinfo{author}{\bibfnamefont{Y.}~\bibnamefont{Sidis}},
  \bibinfo{author}{\bibfnamefont{P.}~\bibnamefont{Bourges}},
  \bibinfo{author}{\bibfnamefont{V.}~\bibnamefont{Hinkov}},
  \bibinfo{author}{\bibfnamefont{A.}~\bibnamefont{Ivanov}},
  \bibinfo{author}{\bibfnamefont{C.}~\bibnamefont{Ulrich}},
  \bibinfo{author}{\bibfnamefont{L.~P.} \bibnamefont{Regnault}},
  \bibnamefont{and} \bibinfo{author}{\bibfnamefont{B.}~\bibnamefont{Keimer}}
  (\bibinfo{year}{2004}), \eprint{arXiv:cond-mat/0403609}.

\bibitem[{\citenamefont{Tranquada et~al.}(2004)\citenamefont{Tranquada, Woo,
  Perring, Goka, Gu, Xu, Fujita, and Yamada}}]{tranquada04}
\bibinfo{author}{\bibfnamefont{J.~M.} \bibnamefont{Tranquada}},
  \bibinfo{author}{\bibfnamefont{H.}~\bibnamefont{Woo}},
  \bibinfo{author}{\bibfnamefont{T.~G.} \bibnamefont{Perring}},
  \bibinfo{author}{\bibfnamefont{H.}~\bibnamefont{Goka}},
  \bibinfo{author}{\bibfnamefont{G.~D.} \bibnamefont{Gu}},
  \bibinfo{author}{\bibfnamefont{G.}~\bibnamefont{Xu}},
  \bibinfo{author}{\bibfnamefont{M.}~\bibnamefont{Fujita}}, \bibnamefont{and}
  \bibinfo{author}{\bibfnamefont{K.}~\bibnamefont{Yamada}},
  \bibinfo{journal}{Nature} \textbf{\bibinfo{volume}{429}},
  \bibinfo{pages}{534} (\bibinfo{year}{2004}).

\bibitem[{\citenamefont{Hayden et~al.}(2004)\citenamefont{Hayden, Mook, Dai,
  Perring, and Do\u{g}an}}]{hayden04}
\bibinfo{author}{\bibfnamefont{S.~M.} \bibnamefont{Hayden}},
  \bibinfo{author}{\bibfnamefont{H.~A.} \bibnamefont{Mook}},
  \bibinfo{author}{\bibfnamefont{P.}~\bibnamefont{Dai}},
  \bibinfo{author}{\bibfnamefont{T.~G.} \bibnamefont{Perring}},
  \bibnamefont{and}
  \bibinfo{author}{\bibfnamefont{F.}~\bibnamefont{Do\u{g}an}},
  \bibinfo{journal}{Nature} \textbf{\bibinfo{volume}{429}},
  \bibinfo{pages}{531} (\bibinfo{year}{2004}).

\bibitem[{\citenamefont{Christensen et~al.}(2004)\citenamefont{Christensen,
  McMorrow, R{\o}nnow, Lake, Hayden, Aeppli, Perring, Mangkorntong, Nohara, and
  Takagi}}]{christensen04}
\bibinfo{author}{\bibfnamefont{N.~B.} \bibnamefont{Christensen}},
  \bibinfo{author}{\bibfnamefont{D.~F.} \bibnamefont{McMorrow}},
  \bibinfo{author}{\bibfnamefont{H.~M.} \bibnamefont{R{\o}nnow}},
  \bibinfo{author}{\bibfnamefont{B.}~\bibnamefont{Lake}},
  \bibinfo{author}{\bibfnamefont{S.~M.} \bibnamefont{Hayden}},
  \bibinfo{author}{\bibfnamefont{G.}~\bibnamefont{Aeppli}},
  \bibinfo{author}{\bibfnamefont{T.~G.} \bibnamefont{Perring}},
  \bibinfo{author}{\bibfnamefont{M.}~\bibnamefont{Mangkorntong}},
  \bibinfo{author}{\bibfnamefont{M.}~\bibnamefont{Nohara}}, \bibnamefont{and}
  \bibinfo{author}{\bibfnamefont{H.}~\bibnamefont{Takagi}}
  (\bibinfo{year}{2004}), \eprint{arXiv:cond-mat/0403439}.

\bibitem[{\citenamefont{Stock et~al.}(2004{\natexlab{a}})\citenamefont{Stock,
  Buyers, Liang, Peets, Tun, Bonn, Hardy, and Birgeneau}}]{stock04_prb}
\bibinfo{author}{\bibfnamefont{C.}~\bibnamefont{Stock}},
  \bibinfo{author}{\bibfnamefont{W.~J.~L.} \bibnamefont{Buyers}},
  \bibinfo{author}{\bibfnamefont{R.}~\bibnamefont{Liang}},
  \bibinfo{author}{\bibfnamefont{D.}~\bibnamefont{Peets}},
  \bibinfo{author}{\bibfnamefont{Z.}~\bibnamefont{Tun}},
  \bibinfo{author}{\bibfnamefont{D.}~\bibnamefont{Bonn}},
  \bibinfo{author}{\bibfnamefont{W.~N.} \bibnamefont{Hardy}}, \bibnamefont{and}
  \bibinfo{author}{\bibfnamefont{R.~J.} \bibnamefont{Birgeneau}},
  \bibinfo{journal}{Phys.\ Rev.\ B} \textbf{\bibinfo{volume}{69}},
  \bibinfo{pages}{014502} (\bibinfo{year}{2004}{\natexlab{a}}).

\bibitem[{\citenamefont{Stock et~al.}(2004{\natexlab{b}})\citenamefont{Stock,
  Buyers, Cowley, Clegg, Coldea, Frost, Liang, Peets, Bonn, Hardy, and Birgeneau}}]{stock04}
\bibinfo{author}{\bibfnamefont{C.}~\bibnamefont{Stock}},
  \bibinfo{author}{\bibfnamefont{W.~J.~L.} \bibnamefont{Buyers}},
  \bibinfo{author}{\bibfnamefont{R.~A.} \bibnamefont{Cowley}},
  \bibinfo{author}{\bibfnamefont{P.~S.} \bibnamefont{Clegg}},
  \bibinfo{author}{\bibfnamefont{R.}~\bibnamefont{Coldea}},
  \bibinfo{author}{\bibfnamefont{C.~D.} \bibnamefont{Frost}},
  \bibinfo{author}{\bibfnamefont{R.}~\bibnamefont{Liang}},
  \bibinfo{author}{\bibfnamefont{D.}~\bibnamefont{Peets}},
  \bibinfo{author}{\bibfnamefont{D.}~\bibnamefont{Bonn}},
  \bibinfo{author}{\bibfnamefont{W.~N.} \bibnamefont{Hardy}}, \bibnamefont{and}
  \bibinfo{author}{\bibfnamefont{R.~J.} \bibnamefont{Birgeneau}},
  (\bibinfo{year}{2004}{\natexlab{b}}),
  \eprint{arXiv:cond-mat/0408071}.

\bibitem[{\citenamefont{Lake et~al.}(2001)\citenamefont{Lake, Aeppli, Clausen,
  McMorrow, Lefmann, Hussey, Mangkorntong, Nohara, Takagi, Mason
  et~al.}}]{lake01}
\bibinfo{author}{\bibfnamefont{B.}~\bibnamefont{Lake}},
  \bibinfo{author}{\bibfnamefont{G.}~\bibnamefont{Aeppli}},
  \bibinfo{author}{\bibfnamefont{K.~N.} \bibnamefont{Clausen}},
  \bibinfo{author}{\bibfnamefont{D.~F.} \bibnamefont{McMorrow}},
  \bibinfo{author}{\bibfnamefont{K.}~\bibnamefont{Lefmann}},
  \bibinfo{author}{\bibfnamefont{N.~E.} \bibnamefont{Hussey}},
  \bibinfo{author}{\bibfnamefont{N.}~\bibnamefont{Mangkorntong}},
  \bibinfo{author}{\bibfnamefont{M.}~\bibnamefont{Nohara}},
  \bibinfo{author}{\bibfnamefont{H.}~\bibnamefont{Takagi}},
  \bibinfo{author}{\bibfnamefont{T.~E.} \bibnamefont{Mason}},
  \bibnamefont{et~al.}, \bibinfo{journal}{Science}
  \textbf{\bibinfo{volume}{291}}, \bibinfo{pages}{1759} (\bibinfo{year}{2001}).

\bibitem[{\citenamefont{Miller et~al.}(2002)\citenamefont{Miller, Kiefl,
  Brewer, Sonier, Chakhalian, Dunsiger, Morris, Price, Bonn, Hardy
  et~al.}}]{miller02}
\bibinfo{author}{\bibfnamefont{R.~I.} \bibnamefont{Miller}},
  \bibinfo{author}{\bibfnamefont{R.~F.} \bibnamefont{Kiefl}},
  \bibinfo{author}{\bibfnamefont{J.~H.} \bibnamefont{Brewer}},
  \bibinfo{author}{\bibfnamefont{J.~E.} \bibnamefont{Sonier}},
  \bibinfo{author}{\bibfnamefont{J.}~\bibnamefont{Chakhalian}},
  \bibinfo{author}{\bibfnamefont{S.}~\bibnamefont{Dunsiger}},
  \bibinfo{author}{\bibfnamefont{G.~D.} \bibnamefont{Morris}},
  \bibinfo{author}{\bibfnamefont{A.~N.} \bibnamefont{Price}},
  \bibinfo{author}{\bibfnamefont{D.~A.} \bibnamefont{Bonn}},
  \bibinfo{author}{\bibfnamefont{W.~H.} \bibnamefont{Hardy}},
  \bibnamefont{et~al.}, \bibinfo{journal}{Phys.\ Rev.\ Lett.}
  \textbf{\bibinfo{volume}{88}}, \bibinfo{pages}{137002}
  (\bibinfo{year}{2002}).

\bibitem[{\citenamefont{Lake et~al.}(2002)\citenamefont{Lake, R{\o}nnow,
  Christensen, Aeppli, Lefmann, McMorrow, Vorderwisch, Smeibidl, Mangkorntong,
  Sasagawa et~al.}}]{lake02}
\bibinfo{author}{\bibfnamefont{B.}~\bibnamefont{Lake}},
  \bibinfo{author}{\bibfnamefont{H.~M.} \bibnamefont{R{\o}nnow}},
  \bibinfo{author}{\bibfnamefont{N.~B.} \bibnamefont{Christensen}},
  \bibinfo{author}{\bibfnamefont{G.}~\bibnamefont{Aeppli}},
  \bibinfo{author}{\bibfnamefont{K.}~\bibnamefont{Lefmann}},
  \bibinfo{author}{\bibfnamefont{D.~F.} \bibnamefont{McMorrow}},
  \bibinfo{author}{\bibfnamefont{P.}~\bibnamefont{Vorderwisch}},
  \bibinfo{author}{\bibfnamefont{P.}~\bibnamefont{Smeibidl}},
  \bibinfo{author}{\bibfnamefont{N.}~\bibnamefont{Mangkorntong}},
  \bibinfo{author}{\bibfnamefont{T.}~\bibnamefont{Sasagawa}},
  \bibnamefont{et~al.}, \bibinfo{journal}{Nature}
  \textbf{\bibinfo{volume}{415}}, \bibinfo{pages}{299} (\bibinfo{year}{2002}).

\bibitem[{\citenamefont{Katano et~al.}(2000)\citenamefont{Katano, Sato, Yamada,
  Suzuki, and Fukase}}]{katano00}
\bibinfo{author}{\bibfnamefont{S.}~\bibnamefont{Katano}},
  \bibinfo{author}{\bibfnamefont{M.}~\bibnamefont{Sato}},
  \bibinfo{author}{\bibfnamefont{K.}~\bibnamefont{Yamada}},
  \bibinfo{author}{\bibfnamefont{T.}~\bibnamefont{Suzuki}}, \bibnamefont{and}
  \bibinfo{author}{\bibfnamefont{T.}~\bibnamefont{Fukase}},
  \bibinfo{journal}{Phys.\ Rev.\ B} \textbf{\bibinfo{volume}{62}},
  \bibinfo{pages}{R14677} (\bibinfo{year}{2000}).

\bibitem[{\citenamefont{Khaykovich et~al.}(2002)\citenamefont{Khaykovich, Lee,
  Erwin, Lee, Wakimoto, Thomas, Kastner, and Birgeneau}}]{khaykovich02}
\bibinfo{author}{\bibfnamefont{B.}~\bibnamefont{Khaykovich}},
  \bibinfo{author}{\bibfnamefont{Y.~S.} \bibnamefont{Lee}},
  \bibinfo{author}{\bibfnamefont{R.} \bibnamefont{Erwin}},
  \bibinfo{author}{\bibfnamefont{S.-H.} \bibnamefont{Lee}},
  \bibinfo{author}{\bibfnamefont{S.}~\bibnamefont{Wakimoto}},
  \bibinfo{author}{\bibfnamefont{K.~J.} \bibnamefont{Thomas}},
  \bibinfo{author}{\bibfnamefont{M.~A.} \bibnamefont{Kastner}},
  \bibnamefont{and} \bibinfo{author}{\bibfnamefont{R.~J.}
  \bibnamefont{Birgeneau}}, \bibinfo{journal}{Phys.\ Rev.\ B}
  \textbf{\bibinfo{volume}{66}}, \bibinfo{pages}{014528}
  (\bibinfo{year}{2002}).

\bibitem[{\citenamefont{Arovas et~al.}(1997)\citenamefont{Arovas, Berlinsky,
  Kallin, and Zhang}}]{arovas97}
\bibinfo{author}{\bibfnamefont{D.~P.} \bibnamefont{Arovas}},
  \bibinfo{author}{\bibfnamefont{A.~J.} \bibnamefont{Berlinsky}},
  \bibinfo{author}{\bibfnamefont{C.}~\bibnamefont{Kallin}}, \bibnamefont{and}
  \bibinfo{author}{\bibfnamefont{S.-C.} \bibnamefont{Zhang}},
  \bibinfo{journal}{Phys.\ Rev.\ Lett.} \textbf{\bibinfo{volume}{79}},
  \bibinfo{pages}{2871} (\bibinfo{year}{1997}).

\bibitem[{\citenamefont{Demler et~al.}(2001)\citenamefont{Demler, Sachdev, and
  Zhang}}]{sachdev01}
\bibinfo{author}{\bibfnamefont{E.}~\bibnamefont{Demler}},
  \bibinfo{author}{\bibfnamefont{S.}~\bibnamefont{Sachdev}}, \bibnamefont{and}
  \bibinfo{author}{\bibfnamefont{Y.}~\bibnamefont{Zhang}},
  \bibinfo{journal}{Phys.\ Rev.\ Lett.} \textbf{\bibinfo{volume}{87}},
  \bibinfo{pages}{067202} (\bibinfo{year}{2001}).

\bibitem[{\citenamefont{Zhang et~al.}(2002)\citenamefont{Zhang, Demler, and
  Sachdev}}]{sachdev02}
\bibinfo{author}{\bibfnamefont{Y.}~\bibnamefont{Zhang}},
  \bibinfo{author}{\bibfnamefont{E.}~\bibnamefont{Demler}}, \bibnamefont{and}
  \bibinfo{author}{\bibfnamefont{S.}~\bibnamefont{Sachdev}},
  \bibinfo{journal}{Phys.\ Rev.\ B} \textbf{\bibinfo{volume}{66}},
  \bibinfo{pages}{094501} (\bibinfo{year}{2002}).

\bibitem[{\citenamefont{Sachdev and Zhang}(2002)}]{sachdev02_science}
\bibinfo{author}{\bibfnamefont{S.}~\bibnamefont{Sachdev}} \bibnamefont{and}
  \bibinfo{author}{\bibfnamefont{S.-C.} \bibnamefont{Zhang}},
  \bibinfo{journal}{Science} \textbf{\bibinfo{volume}{295}},
  \bibinfo{pages}{452} (\bibinfo{year}{2002}).

\bibitem[{\citenamefont{Ghosal et~al.}(2002)\citenamefont{Ghosal, Kallin, and
  Berlinsky}}]{ghosal02}
\bibinfo{author}{\bibfnamefont{A.}~\bibnamefont{Ghosal}},
  \bibinfo{author}{\bibfnamefont{C.}~\bibnamefont{Kallin}}, \bibnamefont{and}
  \bibinfo{author}{\bibfnamefont{A.~J.} \bibnamefont{Berlinsky}},
  \bibinfo{journal}{Phys.\ Rev.\ B} \textbf{\bibinfo{volume}{66}},
  \bibinfo{pages}{214502} (\bibinfo{year}{2002}).

\bibitem[{\citenamefont{Chen et~al.}(2002{\natexlab{b}})\citenamefont{Chen,
  Wang, Zhu, and Ting}}]{ting02_prl}
\bibinfo{author}{\bibfnamefont{Y.}~\bibnamefont{Chen}},
  \bibinfo{author}{\bibfnamefont{Z.~D.} \bibnamefont{Wang}},
  \bibinfo{author}{\bibfnamefont{J.-X.} \bibnamefont{Zhu}}, \bibnamefont{and}
  \bibinfo{author}{\bibfnamefont{C.~S.} \bibnamefont{Ting}},
  \bibinfo{journal}{Phys.\ Rev.\ Lett.} \textbf{\bibinfo{volume}{89}},
  \bibinfo{pages}{217001} (\bibinfo{year}{2002}{\natexlab{b}}).

\bibitem[{\citenamefont{Zhu et~al.}(2002)\citenamefont{Zhu, Martin, and
  Bishop}}]{zhu02}
\bibinfo{author}{\bibfnamefont{J.-X.} \bibnamefont{Zhu}},
  \bibinfo{author}{\bibfnamefont{I.}~\bibnamefont{Martin}}, \bibnamefont{and}
  \bibinfo{author}{\bibfnamefont{A.~R.} \bibnamefont{Bishop}},
  \bibinfo{journal}{Phys.\ Rev.\ Lett.} \textbf{\bibinfo{volume}{89}},
  \bibinfo{pages}{067003} (\bibinfo{year}{2002}).

\bibitem[{\citenamefont{Khomskii and Freimuth}(1995)}]{khomskii95}
\bibinfo{author}{\bibfnamefont{D.~I.} \bibnamefont{Khomskii}} \bibnamefont{and}
  \bibinfo{author}{\bibfnamefont{A.}~\bibnamefont{Freimuth}},
  \bibinfo{journal}{Phys.\ Rev.\ Lett.} \textbf{\bibinfo{volume}{75}},
  \bibinfo{pages}{1384} (\bibinfo{year}{1995}).

\bibitem[{\citenamefont{Blatter et~al.}(1996)\citenamefont{Blatter, Feigel'man,
  Geshkenbein, Larkin, and van Otterlo}}]{blatter96}
\bibinfo{author}{\bibfnamefont{G.}~\bibnamefont{Blatter}},
  \bibinfo{author}{\bibfnamefont{M.}~\bibnamefont{Feigel'man}},
  \bibinfo{author}{\bibfnamefont{V.}~\bibnamefont{Geshkenbein}},
  \bibinfo{author}{\bibfnamefont{A.}~\bibnamefont{Larkin}}, \bibnamefont{and}
  \bibinfo{author}{\bibfnamefont{A.}~\bibnamefont{van Otterlo}},
  \bibinfo{journal}{Phys.\ Rev.\ Lett.} \textbf{\bibinfo{volume}{77}},
  \bibinfo{pages}{566} (\bibinfo{year}{1996}).

\bibitem[{\citenamefont{Arrigoni et~al.}(2002)\citenamefont{Arrigoni, Harju,
  Hanke, Brendel, and Kivelson}}]{arrigoni02}
\bibinfo{author}{\bibfnamefont{E.}~\bibnamefont{Arrigoni}},
  \bibinfo{author}{\bibfnamefont{A.~P.} \bibnamefont{Harju}},
  \bibinfo{author}{\bibfnamefont{W.}~\bibnamefont{Hanke}},
  \bibinfo{author}{\bibfnamefont{B.}~\bibnamefont{Brendel}}, \bibnamefont{and}
  \bibinfo{author}{\bibfnamefont{S.~A.} \bibnamefont{Kivelson}},
  \bibinfo{journal}{Phys.\ Rev.\ B} \textbf{\bibinfo{volume}{65}},
  \bibinfo{pages}{134503} (\bibinfo{year}{2002}).

\bibitem[{\citenamefont{Gilardi et~al.}(2002)\citenamefont{Gilardi, Mesot,
  Drew, Divakar, Lee, Forgan, Zaharko, Conder, Aswal, Dewhurst
  et~al.}}]{gilardi02}
\bibinfo{author}{\bibfnamefont{R.}~\bibnamefont{Gilardi}},
  \bibinfo{author}{\bibfnamefont{J.}~\bibnamefont{Mesot}},
  \bibinfo{author}{\bibfnamefont{A.}~\bibnamefont{Drew}},
  \bibinfo{author}{\bibfnamefont{U.}~\bibnamefont{Divakar}},
  \bibinfo{author}{\bibfnamefont{S.~L.} \bibnamefont{Lee}},
  \bibinfo{author}{\bibfnamefont{E.~M.} \bibnamefont{Forgan}},
  \bibinfo{author}{\bibfnamefont{O.}~\bibnamefont{Zaharko}},
  \bibinfo{author}{\bibfnamefont{K.}~\bibnamefont{Conder}},
  \bibinfo{author}{\bibfnamefont{V.~K.} \bibnamefont{Aswal}},
  \bibinfo{author}{\bibfnamefont{C.~D.} \bibnamefont{Dewhurst}},
  \bibnamefont{et~al.}, \bibinfo{journal}{Phys.\ Rev.\ Lett.}
  \textbf{\bibinfo{volume}{88}}, \bibinfo{pages}{217003}
  (\bibinfo{year}{2002}).

\bibitem[{\citenamefont{Brown et~al.}(2004)\citenamefont{Brown, Charalambous,
  Jones, Forgan, Kealey, Erb, and Kohlbrecher}}]{brown04}
\bibinfo{author}{\bibfnamefont{S.~P.} \bibnamefont{Brown}},
  \bibinfo{author}{\bibfnamefont{D.}~\bibnamefont{Charalambous}},
  \bibinfo{author}{\bibfnamefont{E.~C.} \bibnamefont{Jones}},
  \bibinfo{author}{\bibfnamefont{E.~M.} \bibnamefont{Forgan}},
  \bibinfo{author}{\bibfnamefont{P.~G.} \bibnamefont{Kealey}},
  \bibinfo{author}{\bibfnamefont{A.}~\bibnamefont{Erb}}, \bibnamefont{and}
  \bibinfo{author}{\bibfnamefont{J.}~\bibnamefont{Kohlbrecher}},
  \bibinfo{journal}{Phys.\ Rev.\ Lett.} \textbf{\bibinfo{volume}{92}},
  \bibinfo{pages}{067004} (\bibinfo{year}{2004}).

\bibitem[{\citenamefont{de~Gennes}(1966)}]{degennes66}
\bibinfo{author}{\bibfnamefont{P.~G.} \bibnamefont{de~Gennes}},
  \emph{\bibinfo{title}{Superconductivity of Metals and Alloys}}
  (\bibinfo{publisher}{Addison-Wesley Publishing Company},
  \bibinfo{address}{New York}, \bibinfo{year}{1966}).

\bibitem[{\citenamefont{Franz et~al.}(2002)\citenamefont{Franz, Sheehy, and
  Te\u{s}anovi\'{c}}}]{franz02}
\bibinfo{author}{\bibfnamefont{M.}~\bibnamefont{Franz}},
  \bibinfo{author}{\bibfnamefont{D.~E.} \bibnamefont{Sheehy}},
  \bibnamefont{and}
  \bibinfo{author}{\bibfnamefont{Z.}~\bibnamefont{Te\u{s}anovi\'{c}}},
  \bibinfo{journal}{Phys.\ Rev.\ Lett.} \textbf{\bibinfo{volume}{88}},
  \bibinfo{pages}{257005} (\bibinfo{year}{2002}).

\bibitem[{\citenamefont{Chen et~al.}(2002{\natexlab{c}})\citenamefont{Chen,
  Chen, and Ting}}]{ting02}
\bibinfo{author}{\bibfnamefont{Y.}~\bibnamefont{Chen}},
  \bibinfo{author}{\bibfnamefont{H.~Y.} \bibnamefont{Chen}}, \bibnamefont{and}
  \bibinfo{author}{\bibfnamefont{C.~S.} \bibnamefont{Ting}},
  \bibinfo{journal}{Phys.\ Rev.\ B} \textbf{\bibinfo{volume}{66}},
  \bibinfo{pages}{104501} (\bibinfo{year}{2002}{\natexlab{c}}).

\bibitem[{\citenamefont{Zhang}(1997)}]{zhang97}
\bibinfo{author}{\bibfnamefont{S.-C.} \bibnamefont{Zhang}},
  \bibinfo{journal}{Science} \textbf{\bibinfo{volume}{275}},
  \bibinfo{pages}{1089} (\bibinfo{year}{1997}).

\bibitem[{\citenamefont{Murakami and Fukuyama}(1998)}]{murakami98}
\bibinfo{author}{\bibfnamefont{M.}~\bibnamefont{Murakami}} \bibnamefont{and}
  \bibinfo{author}{\bibfnamefont{H.}~\bibnamefont{Fukuyama}},
  \bibinfo{journal}{J.\ Phys.\ Soc.\ Jpn.} \textbf{\bibinfo{volume}{67}},
  \bibinfo{pages}{2784} (\bibinfo{year}{1998}).

\bibitem[{\citenamefont{Kyung}(2000)}]{kyung00}
\bibinfo{author}{\bibfnamefont{B.}~\bibnamefont{Kyung}},
  \bibinfo{journal}{Phys.\ Rev.\ B} \textbf{\bibinfo{volume}{62}},
  \bibinfo{pages}{9083} (\bibinfo{year}{2000}).

\bibitem[{\citenamefont{Takigawa et~al.}(2004)\citenamefont{Takigawa, Ichioka,
  and Machida}}]{ichmata04}
\bibinfo{author}{\bibfnamefont{M.}~\bibnamefont{Takigawa}},
  \bibinfo{author}{\bibfnamefont{M.}~\bibnamefont{Ichioka}}, \bibnamefont{and}
  \bibinfo{author}{\bibfnamefont{K.}~\bibnamefont{Machida}},
  \bibinfo{journal}{J.\ Phys.\ Soc.\ Jpn.} \textbf{\bibinfo{volume}{73}},
  \bibinfo{pages}{450} (\bibinfo{year}{2004}).

\bibitem[{\citenamefont{Veillette et~al.}(1999)\citenamefont{Veillette,
  Bazaliy, Berlinsky, and Kallin}}]{veillette99}
\bibinfo{author}{\bibfnamefont{M.}~\bibnamefont{Veillette}},
  \bibinfo{author}{\bibfnamefont{Y.~B.} \bibnamefont{Bazaliy}},
  \bibinfo{author}{\bibfnamefont{A.~J.} \bibnamefont{Berlinsky}},
  \bibnamefont{and} \bibinfo{author}{\bibfnamefont{C.}~\bibnamefont{Kallin}},
  \bibinfo{journal}{Phys.\ Rev.\ Lett.} \textbf{\bibinfo{volume}{83}},
  \bibinfo{pages}{2413} (\bibinfo{year}{1999}).

\bibitem[{\citenamefont{Wang and MacDonald}(1995)}]{wang95}
\bibinfo{author}{\bibfnamefont{Y.}~\bibnamefont{Wang}} \bibnamefont{and}
  \bibinfo{author}{\bibfnamefont{A.~H.} \bibnamefont{MacDonald}},
  \bibinfo{journal}{Phys.\ Rev.\ B} \textbf{\bibinfo{volume}{52}},
  \bibinfo{pages}{R3876} (\bibinfo{year}{1995}).

\bibitem[{\citenamefont{Andersen et~al.}(2000)\citenamefont{Andersen, Bruus,
  and Hedeg{\aa}rd}}]{andersen00}
\bibinfo{author}{\bibfnamefont{B.~M.} \bibnamefont{Andersen}},
  \bibinfo{author}{\bibfnamefont{H.}~\bibnamefont{Bruus}}, \bibnamefont{and}
  \bibinfo{author}{\bibfnamefont{P.}~\bibnamefont{Hedeg{\aa}rd}},
  \bibinfo{journal}{Phys.\ Rev.\ B} \textbf{\bibinfo{volume}{61}},
  \bibinfo{pages}{6298} (\bibinfo{year}{2000}).

\bibitem[{\citenamefont{Ogata}(1999)}]{ogata99}
\bibinfo{author}{\bibfnamefont{M.}~\bibnamefont{Ogata}},
  \bibinfo{journal}{Int.\ J.\ Mod.\ Phys.\ B} \textbf{\bibinfo{volume}{13}},
  \bibinfo{pages}{3560} (\bibinfo{year}{1999}).

\bibitem[{\citenamefont{Zhu and Ting}(2001)}]{ting01}
\bibinfo{author}{\bibfnamefont{J.-X.} \bibnamefont{Zhu}} \bibnamefont{and}
  \bibinfo{author}{\bibfnamefont{C.~S.} \bibnamefont{Ting}},
  \bibinfo{journal}{Phys.\ Rev.\ Lett.} \textbf{\bibinfo{volume}{87}},
  \bibinfo{pages}{147002} (\bibinfo{year}{2001}).

\bibitem[{\citenamefont{Matsumoto and
  Shiba}(1995{\natexlab{a}})}]{matsumoto95_1}
\bibinfo{author}{\bibfnamefont{M.}~\bibnamefont{Matsumoto}} \bibnamefont{and}
  \bibinfo{author}{\bibfnamefont{H.}~\bibnamefont{Shiba}},
  \bibinfo{journal}{J.\ Phys.\ Soc.\ Jpn.} \textbf{\bibinfo{volume}{64}},
  \bibinfo{pages}{1703} (\bibinfo{year}{1995}{\natexlab{a}}).

\bibitem[{\citenamefont{Matsumoto and
  Shiba}(1995{\natexlab{b}})}]{matsumoto95_2}
\bibinfo{author}{\bibfnamefont{M.}~\bibnamefont{Matsumoto}} \bibnamefont{and}
  \bibinfo{author}{\bibfnamefont{H.}~\bibnamefont{Shiba}},
  \bibinfo{journal}{J.\ Phys.\ Soc.\ Jpn} \textbf{\bibinfo{volume}{64}},
  \bibinfo{pages}{3384} (\bibinfo{year}{1995}{\natexlab{b}}).

\bibitem[{\citenamefont{Sigrist et~al.}(1995)\citenamefont{Sigrist, Bailey, and
  Laughlin}}]{sigrist95}
\bibinfo{author}{\bibfnamefont{M.}~\bibnamefont{Sigrist}},
  \bibinfo{author}{\bibfnamefont{D.~B.} \bibnamefont{Bailey}},
  \bibnamefont{and} \bibinfo{author}{\bibfnamefont{R.~B.}
  \bibnamefont{Laughlin}}, \bibinfo{journal}{Phys.\ Rev.\ Lett.}
  \textbf{\bibinfo{volume}{74}}, \bibinfo{pages}{3249} (\bibinfo{year}{1995}).

\bibitem[{\citenamefont{Rainer et~al.}(1997)\citenamefont{Rainer, Burkhardt,
  Fogelstr\"{o}m, and Sauls}}]{rainer97}
\bibinfo{author}{\bibfnamefont{D.}~\bibnamefont{Rainer}},
  \bibinfo{author}{\bibfnamefont{H.}~\bibnamefont{Burkhardt}},
  \bibinfo{author}{\bibfnamefont{M.}~\bibnamefont{Fogelstr\"{o}m}},
  \bibnamefont{and} \bibinfo{author}{\bibfnamefont{J.~A.} \bibnamefont{Sauls}}
  (\bibinfo{year}{1997}), \eprint{arXiv:cond-mat/9712234}.

\bibitem[{\citenamefont{Fogelstr\"{o}m
  et~al.}(1997)\citenamefont{Fogelstr\"{o}m, Rainer, and Sauls}}]{fogelstrom97}
\bibinfo{author}{\bibfnamefont{M.}~\bibnamefont{Fogelstr\"{o}m}},
  \bibinfo{author}{\bibfnamefont{D.}~\bibnamefont{Rainer}}, \bibnamefont{and}
  \bibinfo{author}{\bibfnamefont{J.~A.} \bibnamefont{Sauls}},
  \bibinfo{journal}{Phys.\ Rev.\ Lett.} \textbf{\bibinfo{volume}{79}},
  \bibinfo{pages}{281} (\bibinfo{year}{1997}).

\bibitem[{\citenamefont{Samara et~al.}(1990)\citenamefont{Samara, Hammetter,
  and Venturini}}]{samara90}
\bibinfo{author}{\bibfnamefont{G.~A.} \bibnamefont{Samara}},
  \bibinfo{author}{\bibfnamefont{W.~F.} \bibnamefont{Hammetter}},
  \bibnamefont{and} \bibinfo{author}{\bibfnamefont{E.~L.}
  \bibnamefont{Venturini}}, \bibinfo{journal}{Phys.\ Rev.\ B}
  \textbf{\bibinfo{volume}{41}}, \bibinfo{pages}{8974} (\bibinfo{year}{1990}).

\bibitem[{\citenamefont{Chen et~al.}(1991)\citenamefont{Chen, Birgeneau,
  Kastner, Preyer, and Thio}}]{chen91}
\bibinfo{author}{\bibfnamefont{C.~Y.} \bibnamefont{Chen}},
  \bibinfo{author}{\bibfnamefont{R.~J.} \bibnamefont{Birgeneau}},
  \bibinfo{author}{\bibfnamefont{M.~A.} \bibnamefont{Kastner}},
  \bibinfo{author}{\bibfnamefont{N.~W.} \bibnamefont{Preyer}},
  \bibnamefont{and} \bibinfo{author}{\bibfnamefont{T.}~\bibnamefont{Thio}},
  \bibinfo{journal}{Phys.\ Rev.\ B} \textbf{\bibinfo{volume}{43}},
  \bibinfo{pages}{392} (\bibinfo{year}{1991}).

\bibitem[{\citenamefont{Varyukhin and Parfenov}(1993)}]{varyukhin93}
\bibinfo{author}{\bibfnamefont{S.~V.} \bibnamefont{Varyukhin}}
  \bibnamefont{and} \bibinfo{author}{\bibfnamefont{O.~E.}
  \bibnamefont{Parfenov}}, \bibinfo{journal}{Pis'ma Zh.\ \'{E}ksp.\ Teor.\ Fiz.}
  \textbf{\bibinfo{volume}{58}}, \bibinfo{pages}{98} (\bibinfo{year}{1993}) [JETP Lett.\ \textbf{58}, 101 (1993)].

\bibitem[{\citenamefont{Kastner et~al.}(1998)\citenamefont{Kastner, Birgeneau,
  Shirane, and Endoh}}]{kastner98}
\bibinfo{author}{\bibfnamefont{M.~A.} \bibnamefont{Kastner}},
  \bibinfo{author}{\bibfnamefont{R.~J.} \bibnamefont{Birgeneau}},
  \bibinfo{author}{\bibfnamefont{G.}~\bibnamefont{Shirane}}, \bibnamefont{and}
  \bibinfo{author}{\bibfnamefont{Y.}~\bibnamefont{Endoh}},
  \bibinfo{journal}{Rev.\ Mod.\ Phys.} \textbf{\bibinfo{volume}{70}},
  \bibinfo{pages}{897} (\bibinfo{year}{1998}).

\bibitem[{\citenamefont{Kumagai et~al.}(2001)\citenamefont{ichi Kumagai,
  Nozaki, and Matsuda}}]{kumagai01}
\bibinfo{author}{\bibfnamefont{K.~I.}~\bibnamefont{Kumagai}},
  \bibinfo{author}{\bibfnamefont{K.}~\bibnamefont{Nozaki}}, \bibnamefont{and}
  \bibinfo{author}{\bibfnamefont{Y.}~\bibnamefont{Matsuda}},
  \bibinfo{journal}{Phys.\ Rev.\ B} \textbf{\bibinfo{volume}{63}},
  \bibinfo{pages}{144502} (\bibinfo{year}{2001}).

\bibitem[{\citenamefont{Nagaoka et~al.}(1998)\citenamefont{Nagaoka, Matsuda,
  Obara, Sawa, Terashima, Chong, Takano, and Suzuki}}]{nagaoka98}
\bibinfo{author}{\bibfnamefont{T.}~\bibnamefont{Nagaoka}},
  \bibinfo{author}{\bibfnamefont{Y.}~\bibnamefont{Matsuda}},
  \bibinfo{author}{\bibfnamefont{H.}~\bibnamefont{Obara}},
  \bibinfo{author}{\bibfnamefont{A.}~\bibnamefont{Sawa}},
  \bibinfo{author}{\bibfnamefont{T.}~\bibnamefont{Terashima}},
  \bibinfo{author}{\bibfnamefont{I.}~\bibnamefont{Chong}},
  \bibinfo{author}{\bibfnamefont{M.}~\bibnamefont{Takano}}, \bibnamefont{and}
  \bibinfo{author}{\bibfnamefont{M.}~\bibnamefont{Suzuki}},
  \bibinfo{journal}{Phys.\ Rev.\ Lett.} \textbf{\bibinfo{volume}{80}},
  \bibinfo{pages}{3594} (\bibinfo{year}{1998}).

\bibitem[{\citenamefont{Clayhold et~al.}(2003)\citenamefont{Clayhold, Fleming,
  and Skove}}]{clayhold03}
\bibinfo{author}{\bibfnamefont{J.~A.} \bibnamefont{Clayhold}},
  \bibinfo{author}{\bibfnamefont{T.~S.} \bibnamefont{Fleming}},
  \bibnamefont{and} \bibinfo{author}{\bibfnamefont{M.~J.} \bibnamefont{Skove}},
  \bibinfo{journal}{Physica C} \textbf{\bibinfo{volume}{391}},
  \bibinfo{pages}{272} (\bibinfo{year}{2003}).

\bibitem[{\citenamefont{Pickett}(1989)}]{pickett}
\bibinfo{author}{\bibfnamefont{W.~E.} \bibnamefont{Pickett}},
  \bibinfo{journal}{Computer Physics Reports} \textbf{\bibinfo{volume}{9}},
  \bibinfo{pages}{115} (\bibinfo{year}{1989}).

\bibitem[{\citenamefont{Raczkowski et~al.}(2001)\citenamefont{Raczkowski,
  Canning, and Wang}}]{raczkowski01}
\bibinfo{author}{\bibfnamefont{D.}~\bibnamefont{Raczkowski}},
  \bibinfo{author}{\bibfnamefont{A.}~\bibnamefont{Canning}}, \bibnamefont{and}
  \bibinfo{author}{\bibfnamefont{L.~W.} \bibnamefont{Wang}},
  \bibinfo{journal}{Phys.\ Rev.\ B} \textbf{\bibinfo{volume}{64}},
  \bibinfo{pages}{121101(R)} (\bibinfo{year}{2001}).

\bibitem[{\citenamefont{Vanderbilt and Louie}(1984)}]{vanderbilt84}
\bibinfo{author}{\bibfnamefont{D.}~\bibnamefont{Vanderbilt}} \bibnamefont{and}
  \bibinfo{author}{\bibfnamefont{S.~G.} \bibnamefont{Louie}},
  \bibinfo{journal}{Phys.\ Rev.\ B} \textbf{\bibinfo{volume}{30}},
  \bibinfo{pages}{6118} (\bibinfo{year}{1984}).

\bibitem[{\citenamefont{Johnson}(1988)}]{johnson}
\bibinfo{author}{\bibfnamefont{D.~D.} \bibnamefont{Johnson}},
  \bibinfo{journal}{Phys.\ Rev.\ B} \textbf{\bibinfo{volume}{38}},
  \bibinfo{pages}{12807} (\bibinfo{year}{1988}).

\bibitem[{\citenamefont{Chen and Ting}(2004)}]{ting04}
\bibinfo{author}{\bibfnamefont{Y.}~\bibnamefont{Chen}} \bibnamefont{and}
  \bibinfo{author}{\bibfnamefont{C.~S.} \bibnamefont{Ting}},
  \bibinfo{journal}{Phys.\ Rev.\ Lett.} \textbf{\bibinfo{volume}{92}},
  \bibinfo{pages}{077203} (\bibinfo{year}{2004}).

\bibitem[{\citenamefont{Atkinson}(2004)}]{atkinson04}
\bibinfo{author}{\bibfnamefont{W.~A.} \bibnamefont{Atkinson}}
  (\bibinfo{year}{2004}), \eprint{arXiv:cond-mat/0407450}.

\end{thebibliography}
\end{document}